# A Nonlocal Damage-enhanced Lattice Particle Model for Ductile Fracture Analysis

*Changyu Meng[a], Yongming Liu[a,*]*

[a] School for Engineering of Matter, Transport and Energy, Arizona State University, Tempe, AZ 85281

**Abstract**

Ductile fracture of metallic materials typically involves the elastoplastic deformation and associated damaging process. The nonlocal lattice particle method (LPM) can be extended to model this complex behavior. Recently, a distortional energy-based model is formulated into LPM to simulate J2 plasticity. However, this model is based on the incremental updating algorithm which needs very small loading steps to get reasonable results. This is time-consuming and unstable for large systems. Therefore, in this paper, a tensor-based return-mapping algorithm is proposed to deal with these deficiencies. The material deterioration process is modelled as a nonlocal damage evolution process. The particle-size/lattice dependency of macroscopic mechanical responses are handled properly by using the proposed model. Numerical examples of predicting the elastoplastic behavior of engineering structures with/without damage and fracture are also provided. A multi-threaded implementation of LPM using C is available on the website: https://github.com/ymlasu/LPM-C.

---

[*] Corresponding author: Y. Liu (Yongming.Liu@asu.edu), Tel.: +1 480 965 6883, Address: 501 E Tyler Mall, ECG 301, Tempe, AZ 85287, USA

# Highlights

- A nonlocal damage-enhanced lattice particle model was proposed for ductile fracture analysis
- The simulation of plastic behavior using return mapping method is stable and general
- Ductile fracture prediction using nonlocal damage converges with small particle size

## Nomenclature

| | |
|---|---|
| $A$ | material parameter in Oyane damage model |
| $C_{IJKL}$ | material stiffness tensor |
| $e$ | error between actual bond strains and theoretical one |
| $D_c$ | critical damage value |
| $D^{ij}$ | damage parameter for bond $ij$ |
| $\dot{D}^{loc}(x_j)$ | local damage rate of particle $j$ |
| $E$ | Young's modulus |
| $f(x_i)$ | probability density function of Gaussian distribution |
| $\boldsymbol{F}_{ij}$ | interaction force (vector) within the bond $ij$ |
| $F_{ij}^D$ | bond force if considering bond damage |
| $\boldsymbol{F}^{in}$ | internal force vector |
| $\boldsymbol{F}^{ex}$ | external forces vector |
| $g$ | isotropic hardening function |
| $H$ | linear isotropic hardening modulus |
| $\boldsymbol{K}$ | global tangent stiffness matrix |
| $\boldsymbol{K}_i$ | local stiffness matrix for particle $i$ |
| $K^c$ | LPM model parameters for local pair-wise interaction of c-th unit cell |
| $l_{ij}^c, \delta l_{ij}^c$ | bond length, bond elongation between particles $i$ and $j$ of c-th unit cell |
| $L$ | characteristic length in nonlocal damage model |
| $n^{AFEM}$ | number of particles in an AFEM element |
| $n^c$ | number of bonds for the $c$-th unit cell |
| $N_p$ | total number of particles in the system |
| $\boldsymbol{N}^{ij}$ | unit bond vector between material particles $i$ and $j$ |
| $P(x)$ | total energy of the current particle system |
| $r$ | particle radius |
| $\boldsymbol{R}$ | residual force vector for the particles system |
| $t^{ij}$ | traction vector along bond $ij$ |
| $T$ | LPM model parameters for nonlocal interaction |
| $\boldsymbol{u}$ | displacement vector of total particle system |
| $\boldsymbol{u}_i$ | displacement vectors for particle $i$ |
| $U, U_{cell}^c$ | potential energy of all unit cells, the $c$-th unit cell |
| $U_{local}^c, U_{nonlocal}^c$ | local, nonlocal energy for the $c$-th unit cell |
| $V^i$ | volume of particle $i$ |
| $w^{ij}$ | weight parameter for damage modelling |
| $\boldsymbol{x}^*, \boldsymbol{x}$ | equilibrium, current position vector of all particles |

A Nonlocal Damage-enhanced Lattice Particle Model for Ductile Fracture Analysis| | |
|---|---|
| $\boldsymbol{x}_k$ | position vector of the particle $k$ |
| $\alpha$ | equivalent plastic strain |
| $\boldsymbol{\beta}$ | back stress tensor |
| $\Gamma$ | deterioration ratio |
| $\delta l_{ij}^{e,c}, \delta l_{ij}^{p,c}$ | elastic and plastic parts of total bond elongations |
| $\Delta$ | disturbance quantity |
| $\boldsymbol{\varepsilon}$ | total strain tensor |
| $\boldsymbol{\varepsilon}^p$ | plastic strain tensor |
| $\Theta(\boldsymbol{x}_i)$ | integral quantity for Gaussian probability density function |
| $^c\kappa_{KL}, \, ^r\kappa_{IJKL}$ | tensors used in weighted least squares method |
| $\dot{\lambda}$ | plasticity multiplier |
| $\xi_{cr}$ | critical bond strain for brittle damage |
| $\xi^{ij}$ | bond strains for the bond $ij$ |
| $\mu$ | shear modulus |
| $\nu$ | Poisson's ratio |
| $\eta$ | mixed hardening indicator |
| $\boldsymbol{\sigma}$ | stress tensor |
| $\sigma_m$ | hydrostatic stress |
| $\sigma_y$ | yield stress |
| $\boldsymbol{\sigma}^d$ | deviatoric stress tensor |
| $\Phi$ | yield function |

# 1. Introduction

Ductile fracture of metallic materials typically involves the failure by micro-void cavitation or by plastic instability [1]. Such a damaging process will gradually deteriorate the load-carrying capacity of materials until totally failed, especially near the region of stress concentration [2]. This phenomenon is highly nonlinear and involves discontinuous crack growth, which indicates that the plasticity, damage, and fracture should be integrated together to study the ductile failure mechanism, as well as the stress state, damage patterns, and the crack propagation behaviors of engineering structures.

Many damage models for ductile fracture phenomena have been proposed. Ambati et al. [3] divided these approaches into two subcategories, i.e., continuum-based models and discontinuous ones. The well-known Gurson–Tvergaard–Needleman (GTN) models [4,5] and continuum damage models (CDM) [6–8] are continuum-based. On the other hand, cohesive zone models (CZM) [9] utilizes discontinuous formulations since interfaces are explicitly expressed in the model. Finite element method (FEM) is usually involved to implement these models.

One of critical issues when using FEM to simulate the fracture behaviors comes from the continuous requirement for governing equations [10]. The field quantities are poorly defined near the crack tip. As a result, additional efforts should be done, such as remeshing, state variables transferring, manual selection of crack grow direction, and augmentation with discontinuity terms [3]. A famous example is the extended finite element method (XFEM) [11]. Another example is combined FEM and CZM where cohesive elements are inserted between finite elements, in which the crack growth sensitivity with mesh shape is unavoidable.

Another major issue in continuum-based models is the mesh size dependence. For example, when using continuum-based GTN or CDM models, the FEM solutions would not converge even though mesh refinement is applied [12]. This is due to the softening and strain localization behavior in damaged regions. One of the solutions to deal with mesh size dependency issues is to consider nonlocality in the model formulation. Spatial interaction terms using non-local damage models [12,13] can be adopted, like the gradient damage model or phase field models (PFM) [14,15]. Another solution is to construct a set of integral or integro-differential equations rather than partial differential equations to describe the continuum mechanics problems [16,17], e.g., peridynamics (PD) methods [16] or nonlocal lattice methods [18]. Lattice based nonlocal formulations is adopted in this work to explicitly represent the fracture process as well as to avoid the crack path sensitivity issues.

The lattice methods date back to Hrennikoff [19] in 1941, who used the lattice model to analyze the elastic deformation of various structures. In practice, there would be severe fixed-Poisson's ratio problems if directly using the local pair-wise interactions in lattice methods [20,21]. Thus, nonlocality should be considered into these models to describe not only the pair-wise potentials of real materials, but also the influence from the surroundings. There are several different forms of nonlocality in lattice models, like those with the bond angle potential [22], 4-D interactions [23], shear springs interactions [24] or volumetric change of unit cells [18]. By constructing an ordered or disordered topological structure mimicking natural materials, lattice methods can simulate mechanical responses of metals or granular media [21]. The nonlocal lattice methods that only consider the spring and volumetric energies are the so-called nonlocal lattice particle methods (LPM) [18]. Compared to other lattice methods, the advantage of LPM is that there is no additional degree of freedom being added in the model formulation. By only introducing the spring bonds, LPM has the potential to handle larger material systems than other lattice methods [21]. Thus, this method is the focus of this paper.

The nonlocal LPM methods have shown the success of simulating the fracture process of brittle or quasi-brittle materials [25]. However, as indicated by Zhao et al. [26], the general lattice methods are not suitable for plastic analysis since the real-world plasticity problems usually involve the relationships between strain tensor and stress tensor. This tensor notation is not widely adopted in lattice methods. The trivial adoption of nonlinear bond strain-bond force constitutive relationships [27] would result in the deformation anisotropy issues [28]. Zhao et al. [26] implemented a modified Drucker-Prager plasticity in the distinct lattice spring model (DLSM). However, they utilized an explicit integration framework to simulate the quasi-static deformation, which results in the inaccurate yielding point prediction.

Recently, a maximum distortion energy criterion has been formulated into the nonlocal LPM to study the J2 plasticity flow [28]. This novel formulation successfully reproduced the plastic deformation under static/cyclic multiaxial loadings. However, there are still some problems in this energy-based plasticity model. One of the issues is that no damage state/variable can be included, thus it is hard for ductile fracture simulation. Since the constitutive relation is modeled using nonlocal lattice potentials, it is naturally to include consistent nonlocal damage variables and models, which has not been investigated in the past. This is one of the major motivations of the proposed study. When material deterioration is involved, the damaged/fractured material particles should localize into some limited regions [29]. The size of this region as well as the macroscopic mechanical responses should converge by using smaller particle size. As stated by Besson [13], a characteristic length scale should be explicitly expressed in the damage model to get a localized damaged region. There are a number of nonlocal damage models to deal with this problem [30,31]. By following the essence of these models, the damage associated with a particle in LPM is assumed to be influenced by all particles surrounding it. This is shown to be able to effectively reduce the particle-size dependence issue. To the best knowledge of authors', there is no similar research about the incorporation of nonlocal damage concept with the nonlocal lattice methods in the open literature.

Another issue is that the existing LPM plasticity model (i.e., [28]) utilized the incremental updating method to update the distortional energy yield surface, and the computational error will accumulate during the deformation. Thus, the loading steps should be kept very small to get an accurate prediction. This is very expensive for large scale simulation and is not appropriate for damage modeling as the accumulated error in the bond breaking process is very large. Thus, alternative efficient solution algorithms for plasticity and damage are another major objective of the proposed study.

The paper is organized as following. In Section 2, a comprehensive overview about the lattice particle model formulation and solution procedure is given. Next in Section 3, the tensor-based return mapping algorithm is proposed by considering the interchangeability between particle-wise and bond-wise properties. Following this, the nonlocal damage-enhanced modification of LPM is proposed in Section 4 to model the brittle/ductile fracture phenomena. Finally, in Section 5, some benchmark examples are examined to show the ability of the newly proposed framework to do the complex elastoplastic and damage/fracture analysis.

## 2. Overview of nonlocal lattice particle model (LPM)

In this section, the formulation of the nonlocal lattice particle model is introduced and the solution procedure is illustrated. In the following, the uppercases of Latin letters, such as I and J, denote the coordinate components, while the lowercases, such as $i$ and $j$, indicate the particle numbers.

### 2.1 Formulation of the nonlocal lattice particle model

Nonlocal lattice particle model (LPM) is described as a group of particles that located at a regular two-/three-dimensional grid [18,32]. LPM is a nonlocal model in the sense that each particle is connected with its neighbors via spring bonds and interacts with each of its neighbors through pair-wise and multibody potentials [18,33]. A material particle can have multiple layers of neighbors, and each has distinct distances from the central particle. Studies have shown that with more layers of neighbors, some of simulation issues like fracture anisotropy might be eliminated [18]. In this work, two layers of neighbors are chosen considering the balance between accuracy and efficiency.

As described above, the potential energy of a material particle consists of the pairwise potential energy of individual bonds and the multi-body potential energy of all the neighboring bonds. The potential energy is calculated additively for each layer of neighbors. Each layer of neighbors is considered as a unit cell. Above all, the total potential energy $U$ can be expressed as

$$U = \sum_{c=1}^{2} U_{cell}^{c} = \sum_{c=1}^{2}(U_{local}^{c} + U_{nonlocal}^{c}), \tag{1}$$

with

$$U_{local}^{c} = \frac{1}{2} K^{c} \sum_{j=1}^{n^{c}} (\delta l_{ij}^{c})^{2}, \tag{1a}$$

$$U_{nonlocal}^{c} = \frac{1}{2} T \left( \sum_{j=1}^{n^{c}} \delta l_{ij}^{c} \right)^{2}, \tag{1b}$$

where $K^c$ and $T$ are the model parameters for a bond connecting material particles $i$ and $j$, $\delta l_{ij}^c$ is the bond elongation, and $n^c$ is the number of bonds for the $c$-th unit cell of particle $i$. The interaction force within the bond $ij$ can be obtained by differentiating the total potential energy of the particle $i$ with respect to the bond elongation as

$$\boldsymbol{F}_{ij} = \frac{\partial U^i}{\partial(\delta l_{ij}^c)} \boldsymbol{N}^{ij} - \frac{\partial U^j}{\partial(\delta l_{ji}^c)} \boldsymbol{N}^{ji} = \left( 2K^c \delta l_{ij}^c + T\left( \sum_{k=1}^{n^c} \delta l_{ik}^c + \sum_{k=1}^{n^c} \delta l_{jk}^c \right) \right) \boldsymbol{N}^{ij}, \tag{2}$$

where $U^i$ is the potential energy of particle $i$, and $\boldsymbol{N}^{ij}$ (and $\boldsymbol{N}^{ji}$) is unit bond vector between material particles $i$ and $j$, which is given for a specific lattice geometry.

The relationship between bond strains $\xi^{ij}$ and particle strain tensor $\varepsilon_{IJ}$ is shown as Eq. (3), which holds true under the assumption of infinitesimal deformation.

$$\xi^{ij} = \frac{\delta l_{ij}^c}{l_{ij}^c} = \boldsymbol{N}^{ij} \cdot \boldsymbol{\varepsilon} \cdot \boldsymbol{N}^{ij} = \varepsilon_{IJ} N_I^{ij} N_J^{ij} \tag{3}$$

Thus, by plugging Eq. (3) into Eq. (1), one can get

$$\begin{aligned} U &= \sum_{c=1}^{2} \left( \frac{1}{2} K^c \sum_{j=1}^{n^c} (\delta l_{ij}^c)^2 + \frac{1}{2} T \left( \sum_{j=1}^{n^c} \delta l_{ij}^c \right)^2 \right) \\ &= \sum_{c=1}^{2} \left( \frac{1}{2} K^c \sum_{j=1}^{n^c} \left( (\delta l_{ij}^c)^2 N_I^{ij} N_J^{ij} N_K^{ij} N_L^{ij} \varepsilon_{IJ} \varepsilon_{KL} \right) + \frac{1}{2} T \left( \sum_{j=1}^{n^c} \delta l_{ij}^c N_K^{ij} N_L^{ij} \varepsilon_{KL} \right)^2 \right) \end{aligned} \tag{4}$$

The material stiffness tensor can then be determined as

$$C_{IJKL} = \frac{1}{V^i} \frac{\partial^2 U}{\partial \varepsilon_{IJ} \partial \varepsilon_{KL}} \tag{5}$$

There are three model parameters that need to be determined in above equations, which are $K^1$, $K^2$ and $T$. By equating the potential energy at a particle to its counterpart of continuum isotropic elastic materials, the relationships between model parameters and Young's modulus $E$ and Poisson's ratio $v$ can be established for 2D plane strain, 2D plane stress and 3D general cases [34,35]. These relationships are listed in detail as Table I and Table II, which are for 2D and 3D lattices respectively. The other model parameters like the bond vectors for different lattice configurations can be found in [18,27] for 2D lattices and [33] for 3D lattices. The $r$ in these tables means particle radius.

Table I. LPM model parameters for 2D lattices

| Lattice | Square | Hexagon |
|---|---|---|
| $n^1$ | 4 | 6 |
| $n^2$ | 4 | 6 |
| $R^1$ | $2r$ | $2r$ |
| $R^2$ | $2\sqrt{2}r$ | $2\sqrt{3}r$ |
| $K^1$ | $\dfrac{E}{2(1+v)}$ | $\dfrac{\sqrt{3}E}{12(1+v)}$ |
| $K^2$ | $\dfrac{E}{4(1+v)}$ | $\dfrac{\sqrt{3}E}{12(1+v)}$ |
| $T$ (plane strain) | $\dfrac{E(1-4v)}{24(2v-1)(v+1)}$ | $\dfrac{\sqrt{3}E(1-4v)}{144(2v-1)(v+1)}$ |
| $T$ (plane stress) | $\dfrac{E(1-3v)}{24(v-1)(v+1)}$ | $\dfrac{\sqrt{3}E(1-3v)}{144(v-1)(v+1)}$ |

Table II. LPM model parameters for 3D lattices

| Lattice | SC | FCC | BCC |
|---|---|---|---|
| $n^1$ | 6 | 12 | 8 |
| $n^2$ | 12 | 6 | 6 |
| $R^1$ | $2r$ | $2r$ | $2r$ |
| $R^2$ | $2\sqrt{2}r$ | $2\sqrt{2}r$ | $\dfrac{4}{3}\sqrt{3}r$ |
| $K^1$ | $\dfrac{rE}{2(1+v)}$ | $\dfrac{\sqrt{2}rE}{2(1+v)}$ | $\dfrac{\sqrt{3}rE}{2(1+v)}$ |
| $K^2$ | $\dfrac{rE}{2(1+v)}$ | $\dfrac{\sqrt{2}rE}{8(1+v)}$ | $\dfrac{\sqrt{3}rE}{3(1+v)}$ |
| $T$ | $\dfrac{rE(1-4v)}{36(2v-1)(v+1)}$ | $\dfrac{\sqrt{2}rE(1-4v)}{48(2v-1)(v+1)}$ | $\dfrac{\sqrt{3}rE(1-4v)}{28(2v-1)(v+1)}$ |

## 2.2 Solution procedure based on energy minimization

Usually when solving the general problems using LPM, an explicit solution framework based on Newton's dynamic equations can be used [27]. This is intuitional and easy to implement. However, this kind of explicit scheme typically requires a very small timestep and is conditionally stable. Its efficiency and accuracy are usually not suitable for solving static or quasi-static problems [36]. In order to address this issue, an implicit solution procedure based on the so-called atomic-scale finite element method (AFEM) [37] is proposed by Lin et al. [38] for 2D lattices and Chen and Liu [33] for 3D lattices. Here we provide a brief summary of this solution procedure.

The basic idea of AFEM is to minimize the total energy to obtain the equilibrium configuration of lattice structures. If there is a disturbance vector $\boldsymbol{u}$ on the equilibrium position $\boldsymbol{x}^*$, the current position vector $\boldsymbol{x}$ for all particles is

$$\boldsymbol{x} = \boldsymbol{x}^* + \boldsymbol{u} \tag{6}$$

After being applied to some external forces, the total energy of the current particle system is

$$P(\boldsymbol{x}) = U(\boldsymbol{x}) - \sum_{i=1}^{N_p} \boldsymbol{F}_i^{ex} \cdot \boldsymbol{u}_i \tag{7}$$

where $N_p$ is the total number of particles in the system, and $\boldsymbol{F}_i^{ex}$ and $\boldsymbol{u}_i$ are the external force and displacement vectors for particle *i* respectively. Then applying the Taylor expansion to the energy near the equilibrium positions and ignoring the higher-order terms, one can get

$$P(\boldsymbol{x}) \approx P(\boldsymbol{x}^*) + \left.\frac{\partial P(\boldsymbol{x})}{\partial \boldsymbol{x}}\right|_{\boldsymbol{x}=\boldsymbol{x}^*} \cdot \boldsymbol{u} + \frac{1}{2}\boldsymbol{u} \cdot \left.\frac{\partial^2 P(\boldsymbol{x})}{\partial \boldsymbol{x} \partial \boldsymbol{x}}\right|_{\boldsymbol{x}=\boldsymbol{x}^*} \cdot \boldsymbol{u} \tag{8}$$

At the state of minimum potential energy, the particle system should satisfy

$$\frac{\partial P(x)}{\partial x} = \mathbf{0} \tag{9}$$

By substituting the Eq. (7) and Eq. (8) into Eq. (9), one can get a linear equation system by rearranging terms as

$$\boldsymbol{Ku} = \boldsymbol{R} \tag{10}$$

where the global tangent stiffness matrix $\boldsymbol{K}$ and residual force $\boldsymbol{R}$ respectively are

$$\boldsymbol{K} = \frac{\partial^2 P}{\partial x \partial x}\bigg|_{x=x^*} = \frac{\partial^2 U}{\partial x \partial x}\bigg|_{x=x^*} = \frac{\partial \boldsymbol{F}^{in}}{\partial x}\bigg|_{x=x^*} \tag{10a}$$

$$\boldsymbol{R} = -\frac{\partial P}{\partial x}\bigg|_{x=x^*} = \boldsymbol{F}^{ex} - \frac{\partial U}{\partial x}\bigg|_{x=x^*} = \boldsymbol{F}^{ex} - \boldsymbol{F}^{in}\big|_{x=x^*} \tag{10b}$$

The calculation of stiffness matrix $\boldsymbol{K}$ is crucial and nontrivial, which involves the evaluation of nonlocal interactions. The internal force vector $\boldsymbol{F}^{in}$ is obtained by computing projections of bond forces along three coordinate axes. Due to the nonlocality and symmetry of bond forces (Eq. (2)), the internal force vector of particle $i$ is not only influenced by its direct neighbors, but also its neighbors' neighbors. Therefore, all particles that have an influence on the central particle should be taken into account when updating bond and internal forces. All surrounding particles that have influences on a material particle's internal force is called an AFEM element. The number of AFEM particles is $n^{AFEM}$. Thus, according to Eq. (10a), the local stiffness matrix for the AFEM element associated with particle $i$ is

$$\boldsymbol{K}_i = \begin{bmatrix} \dfrac{\partial F^{in}_{x_i}}{\partial x_i} & \dfrac{1}{2}\dfrac{\partial F^{in}_{x_i}}{\partial x_1} & \cdots & \dfrac{1}{2}\dfrac{\partial F^{in}_{x_i}}{\partial x_{n^{AFEM}}} \\ \dfrac{1}{2}\dfrac{\partial F^{in}_{x_i}}{\partial x_1} & 0 & \cdots & 0 \\ \vdots & \vdots & \ddots & \vdots \\ \dfrac{1}{2}\dfrac{\partial F^{in}_{x_i}}{\partial x_{n^{AFEM}}} & 0 & \cdots & 0 \end{bmatrix}_{x=x^*} \tag{11}$$

where $\boldsymbol{x}_k$ is the position vector of the particle $k$. In practice, one can compute the element stiffness matrix using finite difference method [39]. In this forward-difference scheme, i.e., Eq. (12) for a typical 2D case, each particle $i$'s internal forces vector is reevaluated by giving some disturbance $\Delta$ to each degree of freedom of a particle $j$ in its AFEM element list. The disturbance is chosen as $10^{-6}r$ in the current work.

$$\frac{\partial F_{x_i}^{in}}{\partial x_j} = \begin{bmatrix} \frac{\partial F_{ix}^{in}}{\partial x_j} & \frac{\partial F_{ix}^{in}}{\partial y_j} \\ \frac{\partial F_{iy}^{in}}{\partial x_j} & \frac{\partial F_{iy}^{in}}{\partial y_j} \end{bmatrix} \approx \begin{bmatrix} \frac{F_{ix}^{in}(x_j+\Delta)-F_{ix}^{in}(x_j)}{\Delta} & \frac{F_{ix}^{in}(y_j+\Delta)-F_{ix}^{in}(y_j)}{\Delta} \\ \frac{F_{iy}^{in}(x_j+\Delta)-F_{iy}^{in}(x_j)}{\Delta} & \frac{F_{iy}^{in}(y_j+\Delta)-F_{iy}^{in}(y_j)}{\Delta} \end{bmatrix} \tag{12}$$

## 3. Improved J2 plasticity model in LPM framework

In a recent paper of Wei et al. [28], a LPM framework was proposed to deal with the J2 plasticity problem based on the idea of separation of dilatational and distortional energy. The solution procedure was based on an elastic predictor/plastic corrector routine. An incremental updating approach was utilized to check and update the yield surface [28]. During each loading step, the incremental bond elongations are first assumed as fully elastic and then distortional bond elongations are determined. Followed by updating the distortional bond force and distortional energy incrementally. If the distortional energy reaches the critical value, all the updating operations in a loading step must be performed a second time. Then given the model parameters at plastic stage (plastic tangent modulus and plastic Poisson's ratio), the displacement field and the quantities mentioned above need to be recalculated. In their work, the idea of back elongation and back bond forces was also proposed to deal with the isotropic/kinematic hardening behavior.

Even though this framework has been successfully applied to simulate the structural responses under uniaxial and multiaxial loadings [28], there are still some deficiencies in it. The most critical one is its loading step dependence. This incremental updating procedure is similar to the explicit integration method used in FEM [10], in which the loading steps should be small to eliminate the accumulating error under (quasi-)static loading. Another drawback is that some features like nonlinear hardening rules, state variables like damage, etc., are hard to be included in this framework. In addition, since the stiffness matrix is updated incrementally after yielding, some behaviors like the perfectly plastic constitutive relationships cannot be simulated because the plastic stiffness matrix would become singular. These issues may be addressed using a tensor-based return mapping framework coupling with iterative solution methods.

## 3.1 Interchangeability between particle-wise and bond-wise properties

To properly model the J2 plasticity in LPM, some tensor quantities, i.e., strain and stress, should be included into the computational framework. However, this cannot be realized without the interchangeability and mapping mechanism between particle-wise (like strain or stress tensors, etc.) and bond-wise properties (bond elongation, bond force, etc.). The first quantity involved is the stress tensor in infinitesimal deformation cases, it is approximated using all the bond forces exerted from a material particle [40], i.e.

$$\sigma_{IJ} = \frac{1}{V^i}\frac{\partial U}{\partial \varepsilon_{IJ}} = \frac{1}{2V^i}\sum_{c=1}^{2}\sum_{j=1}^{n^c} l_{ij}^c F_{ij} N_I^{ij} N_J^{ij}, \tag{13}$$

Along the direction of the bond vector, the traction between particle *i* and *j* is

$$t^{ij} = \boldsymbol{N}^{ij} \cdot \boldsymbol{\sigma} \cdot \boldsymbol{N}^{ij} \tag{14}$$

The bond strains can also be related to the strain tensor $\boldsymbol{\varepsilon}$ at particle *i* under the assumption of infinitesimal deformation, i.e., Eq. (3). However, even though all the bond elongations surrounding a particle *i* are given, the strain tensor is hard to determine. The reason is that the number of neighbors is usually larger than the number of independent components of strain tensor. To solve this overly-constraint problem, the weighted least squares method [40] is utilized. First, the error between actual bond strains and that computed from Eq. (3) is formulated as

$$e = \sum_{c=1}^{2}\sum_{j=1}^{n^c}(\xi^{ij} - \varepsilon_{IJ}N_I^{ij}N_J^{ij})^2 w_j \tag{15}$$

in which $w_j$ is the weight factor associated to neighbor *j*. By minimizing *e* with respect to the strain tensor $\boldsymbol{\varepsilon}$, the following linear equations can be obtained,

$$\varepsilon_{IJ} = {}^c\kappa_{KL} \, {}^r\kappa_{IJKL}^{-1} \tag{16}$$

and,

$${}^c\kappa_{KL} = \sum_{c=1}^{2} \sum_{j=1}^{n^c} \xi^{ij} N_K^{ij} N_L^{ij} w_j \tag{16a}$$

$${}^r\kappa_{IJKL} = \sum_{c=1}^{2} \sum_{j=1}^{n^c} N_I^{ij} N_J^{ij} N_K^{ij} N_L^{ij} w_j \tag{16b}$$

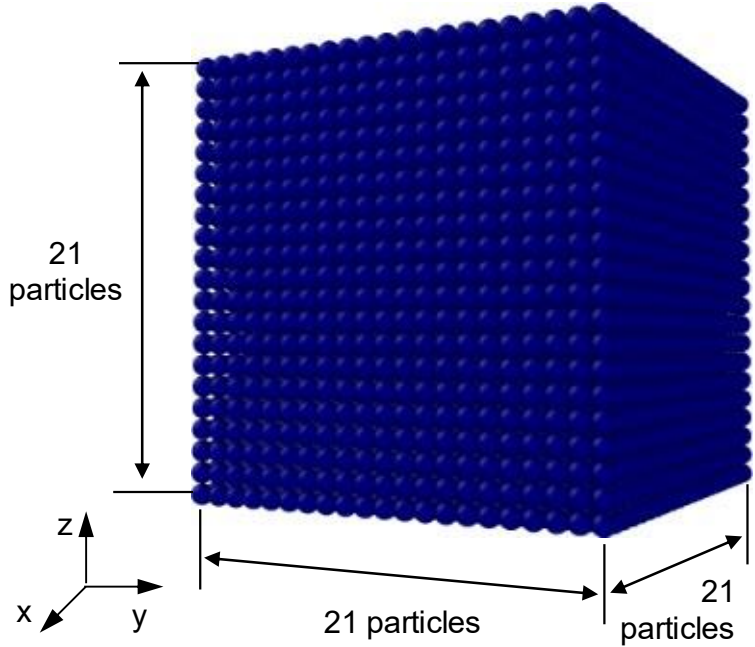

**Figure 1**. Simple cubic lattice system

In the following, a simple cubic lattice shown in Figure 1 is utilized to test the interchangeability between particle-wise and bond-wise properties. The unit bond vectors are listed in Table III for reference. There are 9261 particles in total with each particle's radius being 0.25mm. The lattice system is loaded by distributed force uniaxially along z axis, and the total force is 2000N. The materials constants are listed in the following: the Young's modulus E is 146GPa, Poisson's ratio ν is 0.3. Therefore, the stress and strain states can be determined analytically by the theory of elasticity. After collecting the deformation information of particles with full neighbor list, the

particle-/bond-wise properties can be compared directly.

**Table III**. Bond vectors of simple cubic 3D lattice structure [33]

| Neighbor # | 1 | 2 | 3 | 4 | 5 | 6 |
|---|---|---|---|---|---|---|
| | (1,0,0) | (0,1,0) | (0,0,1) | (-1,0,0) | (0,-1,0) | (0,0,-1) |
| Neighbor # | 7 | 8 | 9 | 10 | 11 | 12 |
| | 1/√2(1,1,0) | 1/√2(1,0,1) | 1/√2(0,1,1) | 1/√2(-1,-1,0) | 1/√2(-1,0,-1) | 1/√2(0,-1,-1) |
| Neighbor # | 13 | 14 | 15 | 16 | 17 | 18 |
| | 1/√2(1,-1,0) | 1/√2(1,0,-1) | 1/√2(0,1,-1) | 1/√2(-1,1,0) | 1/√2(-1,0,1) | 1/√2(0,-1,1) |

In table IV, the continuum physical quantities from both theory and simulation are listed. The quantities with a superscript *theo* are from analytical computation under the given loading condition. Due to the lack of a full neighbor list for particles near the boundary of a simulation domain, the near-surface quantities need some corrections [43] to get reasonable results. In this section, however, such corrections would not be taken into consideration. Thus, the tensor components in Table IV are taken to be the averaged value from particles with full neighbor list. These components are denoted with a superscript *lpm*. To compute the strain tensor, all weight factors $w_j$ are set as 0.5. From the table we can see, both the (theoretical non-zero components) approximated stress and strain tensor values are within less than 5% error compared to the theoretical ones. The simulated values for zero tensor components (such as $\sigma_{11}^{lpm}$, $\varepsilon_{23}^{lpm}$ etc.) are ignorable compared to the non-zero ones. Even though the calculation of strain tensor is an overly-constraint problem, the computational accuracy is still satisfactory.

**Table IV**. Averaged particle-wise information of the lattice system

| Components | 11 | 22 | 33 | 23 | 13 | 12 |
|---|---|---|---|---|---|---|
| $\sigma_{IJ}^{theo}$, MPa | 0.000E+00 | 0.000E+00 | 1.814E+01 | 0.000E+00 | 0.000E+00 | 0.000E+00 |
| $\sigma_{IJ}^{lpm}$, MPa | 1.065E-01 | 1.065E-01 | 1.868E+01 | 1.683E-08 | 5.567E-08 | -1.993E-11 |
| $\varepsilon_{IJ}^{theo}$ | -3.728E-05 | -3.728E-05 | 1.243E-04 | 0.000E+00 | 0.000E+00 | 0.000E+00 |
| $\varepsilon_{IJ}^{lpm}$ | -3.785E-05 | -3.785E-05 | 1.276E-04 | 2.996E-13 | 9.911E-13 | -3.554E-16 |

In table V, on the other hand, we compared the actual bond strains and bond forces with those computed from the mapping from theoretical continuum tensors (ref. Eq. (3) and Eq. (14)). Even though the bond strain values from these two methods match well, the difference between bond force $F_{lpm}^j$ and bond traction $t_{map}^j$ is significant. This indicated that one must include an additional term with the area dimension to convert from force to traction to keep the dimensional unit being consistent. This term is usually lattice-dependent and may vary for different layer of neighbors [18], thus is hard to be applied into real situations.

**Table V**. Averaged bond-wise information of the lattice system

| Neighbor # | 1 | 2 | 3 | 4 | 5 | 6 |
|---|---|---|---|---|---|---|
| $\xi_{map}^j$ | -3.728E-05 | -3.728E-05 | 1.243E-04 | -3.728E-05 | -3.728E-05 | 1.243E-04 |
| $\xi_{lpm}^j$ | -3.785E-05 | -3.785E-05 | 1.276E-04 | -3.785E-05 | -3.785E-05 | 1.276E-04 |
| $t_{map}^j$, MPa | 0.000E+00 | 0.000E+00 | 1.814E+01 | 0.000E+00 | 0.000E+00 | 1.814E+01 |
| $F_{lpm}^j$, N | -4.905E-01 | -4.905E-01 | 1.830E+00 | -4.905E-01 | -4.905E-01 | 1.830E+00 |
| Neighbor # | 7 | 8 | 9 | 10 | 11 | 12 |
| $\xi_{map}^j$ | -3.728E-05 | 4.349E-05 | 4.349E-05 | -3.728E-05 | 4.349E-05 | 4.349E-05 |
| $\xi_{lpm}^j$ | -3.789E-05 | 4.494E-05 | 4.494E-05 | -3.789E-05 | 4.493E-05 | 4.493E-05 |
| $t_{map}^j$, MPa | 0.000E+00 | 9.070E+00 | 9.070E+00 | 0.000E+00 | 9.070E+00 | 9.070E+00 |
| $F_{lpm}^j$, N | -6.382E-01 | 1.004E+00 | 1.004E+00 | -6.382E-01 | 1.004E+00 | 1.004E+00 |
| Neighbor # | 13 | 14 | 15 | 16 | 17 | 18 |
| $\xi_{map}^j$ | -3.728E-05 | 4.349E-05 | 4.349E-05 | -3.728E-05 | 4.349E-05 | 4.349E-05 |
| $\xi_{lpm}^j$ | -3.789E-05 | 4.493E-05 | 4.493E-05 | -3.789E-05 | 4.494E-05 | 4.494E-05 |
| $t_{map}^j$, MPa | 0.000E+00 | 9.070E+00 | 9.070E+00 | 0.000E+00 | 9.070E+00 | 9.070E+00 |
| $F_{lpm}^j$, N | -6.382E-01 | 1.004E+00 | 1.004E+00 | -6.382E-01 | 1.004E+00 | 1.004E+00 |

From the above discussion we can summarize the interchangeability property between bond-wise and particle-wise quantities as shown in Figure 2. It should be noted that the tensors shown in the figure can be generalized as any tensor-form state variables, such as total strain or plastic strain tensors. Therefore, there is one possible routine to update the state variables and internal force states. That is, one starts from bond strain, then bond force, then convert it to stress tensor, then update the state variables using constitutive relationship, finally map to the original bond strain and get the updated bond forces. As indicated in Figure 2, there might be another routine that starts from bond strain to strain tensors. However, this routine will require much more computational time since it involves the solving of linear equations (Eq. (16)). In this work, the first routine is adopted. The detailed instruction of using this routine to iteratively update the state variables will be illustrated in the following sections.

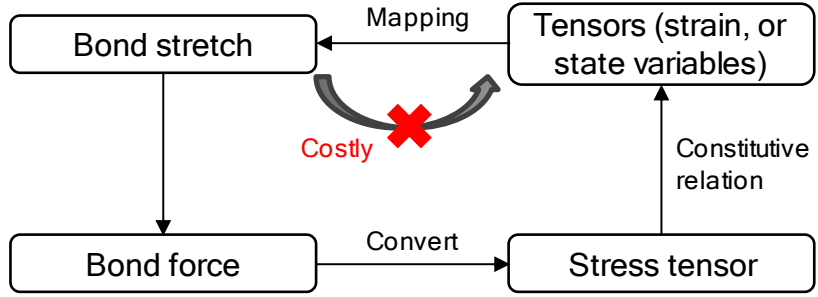

**Figure 2.** Interchangeability property between bond-wise and particle-wise quantities

## 3.2 Improved J2 plasticity formulation using tensor-based constitutive relationships

To address the deficiencies of the energy-based incremental updating algorithm, the return mapping approach using stress tensor is proposed in this section based on the interchangeability properties shown in Figure 2. The continuum plasticity theory is based on the deviatoric stress tensor $\boldsymbol{\sigma}^d$, i.e.

$$\sigma_{IJ}^d = \sigma_{IJ} - \frac{1}{3}\delta_{IJ}\sigma_{IJ} \tag{17}$$

in which $\frac{1}{3}\delta_{IJ}\sigma_{IJ} = \sigma_m$, is the hydrostatic stress. The first ingredient of the J2 plasticity is the yielding function,

$$\Phi = \overline{(\boldsymbol{\sigma}^d - \boldsymbol{\beta})} - g(\alpha) \tag{18}$$

where $\boldsymbol{\beta}$ is the back stress tensor, $g$ is the isotropic hardening rule as a function of equivalent plastic strain $\alpha$. And $\overline{\blacksquare} = \sqrt{\frac{3}{2}\blacksquare:\blacksquare}$ is the J2 norm of the tensor $\blacksquare$. To keep consistent to the mixed linear hardening rule in the work of Wei et al. [28], the linear version of Frederick-Armstrong kinematic hardening [41] and a linear isotropic hardening rule is adopted to form a mixed linear hardening rule, i.e.

$$\begin{cases} \dot{g} = (1-\eta)H\dot{\alpha}, g_0 = \sigma_y \\ \dot{\boldsymbol{\beta}} = \frac{2}{3}\eta H \dot{\boldsymbol{\varepsilon}}^p \quad , \boldsymbol{\beta}_0 = \mathbf{0} \end{cases} \quad (19)$$

in which $H$ is the isotropic hardening modulus, $\eta$ is a mixed hardening indicator that continuously ranges from 0 to 1. $\eta = 0$ means purely isotropic hardening, and $\eta = 1$ means purely linear kinematic hardening. It should be noted that the hardening can be nonlinear in general, and should involve no significant difficulty by adding the nonlinear hardening rule. The plastic flow rule is assumed associative and being formulated using the plastic strain rate tensor and the deviatoric stress tensor [42],

$$\dot{\boldsymbol{\varepsilon}}^p = \dot{\lambda}\frac{\partial \Phi}{\partial \boldsymbol{\sigma}} = \frac{3}{2}\dot{\lambda}\frac{\boldsymbol{\sigma}^d}{\bar{\sigma}^d} \quad (20)$$

The Kuhn-Tucker condition should be satisfied by the system, which is

$$\dot{\lambda} \geq 0, \Phi \leq 0 \text{ and } \dot{\lambda}\Phi = 0 \quad (21)$$

in which $\dot{\lambda}$ is the plasticity multiplier and the Eq. (21) defines the loading/unloading conditions [42]. The tensor-based return mapping framework are based on the elastic predictor-plastic corrector. After the linear system of equations is solved, the trial elastic bond elongation is first computed as following

$$\delta l_{ij,trial}^{e,c} = \delta l_{ij}^c - \delta l_{ij,n}^{p,c} \quad (22)$$

The trial bond force $F_{ij,trial}^c$ and the trial stress tensor $\sigma_{IJ,trial}^i$ would then be computed and then be utilized to construct the trial yield function, i.e., Eq. (23). It should be noted that, the stress tensor might need to be modified near the surface of simulation box since the bonds are incomplete in these regions [43].

$$\Phi_{trial} = \overline{(\boldsymbol{\sigma}_{trial}^d - \boldsymbol{\beta}_n)} - (S_y + (1-\eta) * H\alpha_n) \quad (23)$$

If the yield function $\Phi_{trial}^i > 0$, the plastic multiplier would be larger than zero and need to be determined. At the current timestep, one can get the yield function [41] by combining above equations as

$$\Phi = \overline{(\boldsymbol{\sigma}_{trial}^d - \boldsymbol{\beta}_n)} - \frac{3}{2}\Delta\lambda\left(2\mu + \frac{2}{3}\eta H\right) - (S_y + (1-\eta)*H(\alpha_n + \Delta\lambda)) \tag{24}$$

Combining Eq. (23) with Eq. (24), the incremental plastic multiplier can be solved as

$$\Delta\lambda = \Phi_{trial}/(3\mu + H) \tag{25}$$

For the mixed linear hardening case in this work, the incremental plastic multiplier $\Delta\lambda$ can be calculated analytically without iteration. However, if any of the kinematic or isotropic hardening have nonlinear form, Eq. (24) would likely to be solved iteratively.

Then the state variables can be updated using the newly obtained plastic multiplier, as

$$\Delta\boldsymbol{\varepsilon}^p = \Delta\lambda\frac{3}{2}\frac{\boldsymbol{\sigma}_{trial}^d - \boldsymbol{\beta}_n}{\overline{(\boldsymbol{\sigma}_{trial}^d - \boldsymbol{\beta}_n)}} \tag{26}$$

$$\alpha = \alpha_n + \Delta\lambda \tag{27}$$

$$\boldsymbol{\beta} = \boldsymbol{\beta}_n + \frac{2}{3}\eta H \Delta\boldsymbol{\varepsilon}^p \tag{28}$$

Use the interchangeability property of plastic strain tensor and plastic bond strain, the incremental plastic bond elongation can be obtained as

$$\Delta\delta l_{ij}^{p,c} = l_{ij}^c \Delta\varepsilon_{IJ}^p N_I^{ij} N_J^{ij} \tag{29}$$

$$\delta l_{ij}^{p,c} = \delta l_{ij,n}^{p,c} + \Delta\delta l_{ij}^{p,c} \tag{30}$$

$$\delta l_{ij}^{e,c} = \delta l_{ij}^c - \delta l_{ij}^{p,c} \tag{31}$$

The newly obtained elastic bond elongation would be utilized to update the bond force and stress tensor at particle $i$. As indicated in above, the interchangeability property between plastic strain tensor and plastic bond elongation should be used in the return mapping algorithm for LPM. This is the main difference from the finite element-based method. The computational flowchart of tensor-based return mapping algorithm is summarized in Figure 3.

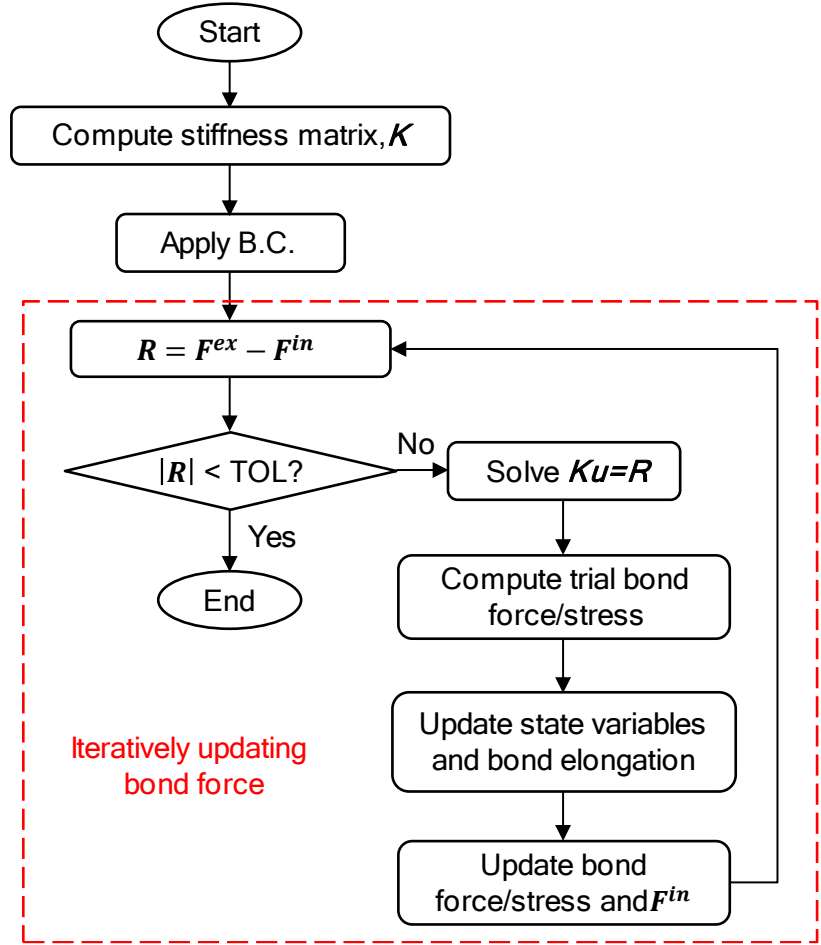

**Figure 3.** Computational flowchart of tensor-based return mapping framework for a typical loading step

## 4. Nonlocal damage-enhanced LPM and fracture modeling

In the nonlocal LPM framework, material failure is typically simulated using the critical bond strain criterion, e.g., [33,44]. Even though this failure criterion is suitable to for brittle fracture problems, it is not appropriate to be used in ductile fracture problems. The reason is that the critical bond strain criterion is similar to the maximum principal stress/strain criterion, which does not take into account other state variables like plastic strain, stress triaxiality ratio, etc. These quantities, however, are shown to be critical in simulating ductile fracture behaviors [7]. For this reason, more general models like the energy-based criterion [45] or the continuum damage mechanics [29] model may be used. Inspired by the work of Tupek et al. [29] and Behzadinasab and Foster [46], the continuum damage mechanics is incorporated into the original LPM model to from a damage-enhanced LPM framework.

In the damage-enhanced lattice particle method, the bond forces are *influenced* by the damage of the bonds between connecting particles.

$$F_{ij}^{\mathrm{D}} = w^{ij} F_{ij} \tag{32}$$

$$w^{ij} = \begin{cases} 1 - D^{ij} & , if\ D^{ij} < D_c \\ 0 & , else \end{cases} \tag{33}$$

in which $w^{ij}$ is the weight parameter that ranges from 0 to 1, and $D^{ij}$ is the damage parameter that keeps tracking the damage status of the current bond. $D_c$ is the critical damage value. $F_{ij}^{\mathrm{D}}$ is the bond force after considering the bond damage. The damage evolution rules may vary for different materials. Bonds will break after the damage reaches to a critical value. It should be noted that, a characteristic length scale should be manually included into the damage formulation. The reason is that the continuous degradation of material would involve softening effects and thus influence the surrounding materials within a certain distance.

## 4.1 Damage modeling

Similar to the critical bond strain criterion used by other lattice particle methods [25], the damage for brittle materials is directly determined by bond strains without the evolution rule. Therefore, the brittle damage associated to bond strain $\xi^{ij}$ is

$$D^{ij} = \begin{cases} 0, & if\ \xi^{ij} \leq \xi_{cr} \\ 1, & else \end{cases} \tag{34}$$

where $\xi_{cr}$ is the critical bond strain. For the ductile fracture simulation, a particular damage accumulation rule should be adopted [47]. However, improper adoption of damage accumulation rule would result in severe lattice dependence issues. As qualitatively analyzed in [13], the dissipated energy is smaller at same crack growth length if using thinner mesh size. This will result in a more brittle-like macroscopic mechanical behavior. The simulation results from the LPM, therefore, would not converge by adopting a smaller particle size. This problem can be addressed by introducing a characteristic length scale into the damage model. Inspired by the work of Tvergaard and Needleman [31] and Enakoutsa et al. [48], the central particle's damage behavior would be influenced by surrounding materials, i.e.

$$\dot{D}(x_i) = \frac{1}{\Theta(x_i)} \int f(x_i - x_j) \dot{D}^{loc}(x_j) dV_j \tag{35}$$

$$\Theta(x_i) = \int f(x_i - x_j) dV_j, f(x_i) = \frac{1}{L\sqrt{2\pi}} exp\left(-\frac{1}{2L^2}\|x_i\|^2\right) \tag{36}$$

in which $f(x_i)$ is the probability density function of Gaussian distribution, and $\dot{D}^{loc}(x_j)$ is the local damage rate of particle $j$. $L$ is a characteristic length scale. It should be noted that the influence from the surrounding particles would decay quickly when the particle distance is increasing. After time-domain integration, the current damage value can be determined as

$$D = \int_0^t \dot{D} d\tau \tag{37}$$

Oyane rule is adopted as the localized damage evolution rule [49], i.e., Eq. (38). This simple damage accumulation rule involves the stress triaxiality $\frac{\sigma^m}{\bar{\sigma}}$ and the equivalent plastic strain rate $\dot{\alpha}$, which is shown to have robust performance to predict the crack initiation location [47]. However, the adoption of more complex local damage evolution rule is also possible.

$$\dot{D}^{loc} = \left(1 + \frac{A\sigma^m}{\bar{\sigma}}\right) \dot{\alpha} \tag{38}$$

in which $A$ is a material parameter that needs to be calibrated. Each time after the particle-wise damage variable is computed globally, each bond's damage can be determined as the maximum damage value of two connected particles. The evolution of damage and integrity variables for brittle and ductile materials are illustrated qualitatively in Figure 4.

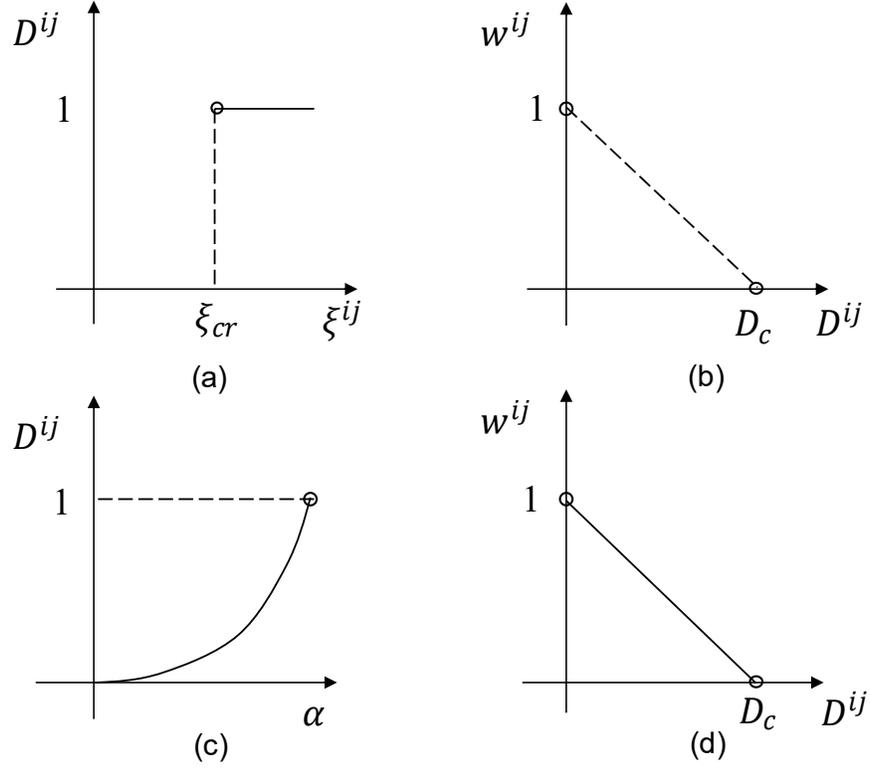

**Figure 4.** Illustration of damage evolution rule and integrity parameter for (a, b) brittle materials and (c, d) ductile materials

The deterioration ratio $\Gamma$ of a particle $i$ is defined as

$$\Gamma = 1 - \frac{\sum_{c=1}^{2}\sum_{j=1}^{n^c} w^{ij}}{\sum_{c=1}^{2} n^c} \tag{39}$$

in which $n^c$ is the number of $i$'s neighboring particles of the c-th unit cell. $\Gamma = 0$ when all bonds are intact, and $\Gamma = 1$ if all bonds connected to particle $i$ are broken.

## 4.2 Solution procedure of nonlocal damage-enhanced LPM with crack propagation

In this section, the global iteration algorithm for elastoplastic problems is modified to couple with crack propagation process. The flowchart is shown in Figure 5. It represents a typical integration step from $t_n$ to $t_{n+1}$ for simulating the brittle/ductile fracture phenomena. As shown in the figure, the constitutive relationship updating procedure is condensed into the update bond force and internal force $\boldsymbol{F}^{in}$ step. If the iteration criterion for minimizing the residual force is reached, the particle-/bond-wise damage have to be updated. Then the topological configuration of the lattice system is modified, as well as the stiffness matrix and internal forces vector.

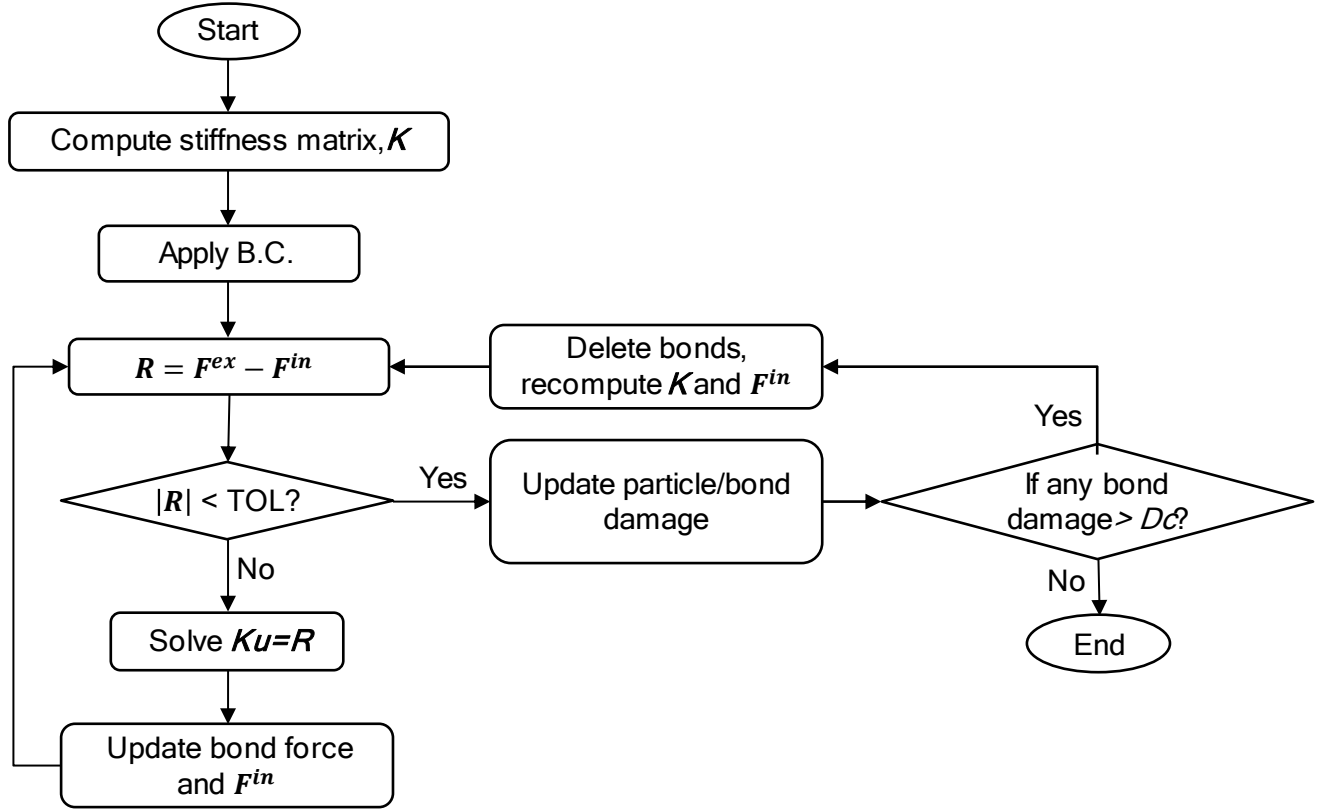

**Figure 5**. Crack propagation algorithm for brittle/ductile fracture simulations

The difference between the current framework and the method adopted by Ni et al. [50] is whether including the iteration procedure. By incorporating the iterative process in the current work, the stress and deformation field would be re-distributed, and the equilibrium can be reached for the entire system. Thus, the equilibrium is guaranteed before moving from one timestep to another.

## 5. Numerical examples

Several benchmark examples are used to demonstrate the proposed damage-enhanced elastoplastic simulation procedure. First, some benchmark examples without damage and fracture are discussed to show the improvement over the original incremental updating algorithm proposed by Wei et al. [28]. Uniaxial monotonic and cyclic loading tests for homogeneous solid cube are included, as well as a 3D plate with a through-thickness central crack. Next, the proposed damage model is incorporated into LPM to simulate the crack propagation behavior. The lattice dependence of brittle crack grow direction and particle size dependence of ductile fracture are investigated. In addition, the simulation results are compared with the experimental fracture phenomena from open literature.

### 5.1 Elastoplastic analysis without damage and fracture
#### *5.1.1 3D cube under uniaxial loadings*

The examples in this subsection show the ability of proposed algorithm to deal with the elastoplastic problems, as well as its advantages over the incremental updating method. The lattice configuration is the same as the one mentioned in Section 3.1, as in Figure 1. The initial yield stress $\sigma_y$ is 200MPa and the hardening modulus $H$ is 38.714MPa.

*Case I: Uniaxial tensile loading*

For both the tensor-based return mapping and the incremental updating methods, 10, 20, and 40 displacement loading steps are prescribed to the system boundary. The total uniaxial strain along z direction ($\varepsilon_{33}$) is 0.2% for all cases. The mixed hardening mode indicator $\eta$ is set as 0 to activate isotropic hardening since only tensile loading is involved in this subsection.

As shown in Figure 6(a), the stress-strain curves using different number of loading steps are identical for return mapping algorithm. However, Wei's incremental updating model shows different behavior and asymptotically approaches the return mapping results after increasing the loading step number. As a result, the proposed method is shown to be more robust for simulation time steps.

Figure 6(b) shows the comparison between theoretical stress-strain curves (along z directions) and the simulating ones using return mapping algorithm with varying particle radius. The mechanical response is more accurate if using smaller particle radius, this is consistent to the conclusion of [18]. However, the simulation time will increase quadratically with decreasing particle radius. On the other hand, the accuracy is not affected much by using a moderate particle radius. Thus, the proposed method also shows robustness with respect to the particle density.

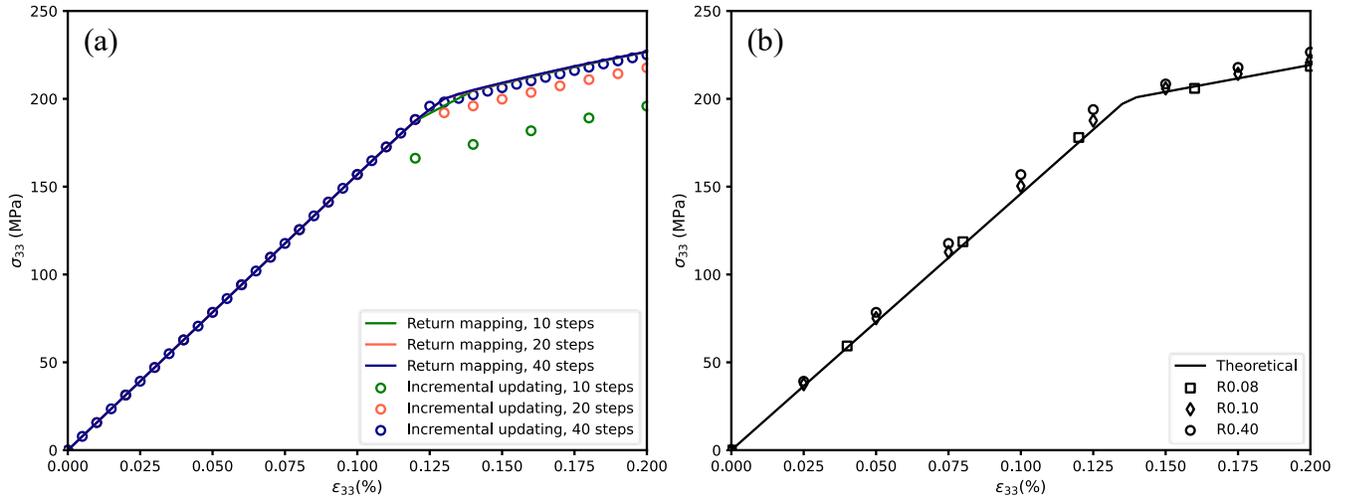

**Figure 6.** a) Comparison of stress-strain curves between incremental updating and tensor-based return mapping algorithms for different loading steps, and b) Computational accuracy of stress-strain curves of return mapping algorithm using different size of particles

*Case II: Uniaxial cyclic loading*

The inaccuracy of incremental updating algorithm would be exaggerated in cyclic loading cases as the simulation error continues to grow cycle by cycle. To study this behavior, the cyclic loading histories are applied to the lattice system, which is shown in Figure 7 (a), (c), and (e). For the simulations using return mapping algorithm, the incremental forces are $\pm 2000$N. However, for incremental updating methods the incremental forces are $\pm 500$N. The hardening mode indicator $\eta$ can be modified from 0 to 1 to test the mixed and kinematic hardening modes.

As shown in Figure 7 (b), there shows significant departures between the stress strain curves of the two methods during the process when loads grow from -40kN to 46kN, even though the elastic and initial isotropic hardening curves in previous steps are almost identical. The inaccuracy of incremental updating method is more apparent in pure kinematic hardening cases, i.e., Figure 7(d), which shows that the stress-strain curve even does not enclose after a full loading cycle. And from Figure 7(b), 7(d), and 7(f) we may conclude that, the incremental updating algorithm always underestimates the yielding point due to the trivial adding of incremental distortional energy.

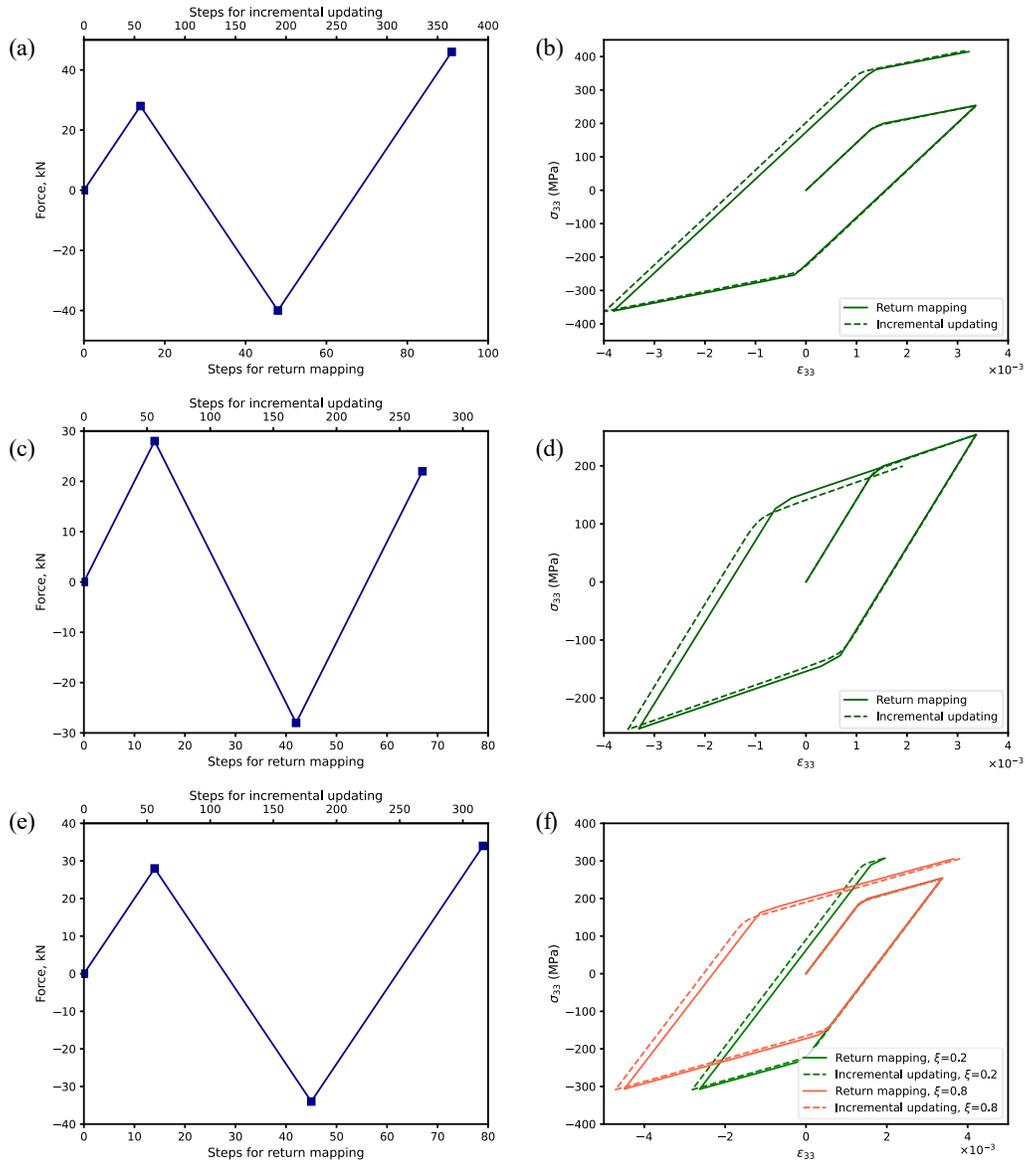

**Figure 7.** Comparison of mechanical responses between return mapping and incremental updating methods. In which a), c), and e) are loading histories; b), d), and f) are stress-strain curves for pure isotropic, pure kinematic, and mixed hardening cases.

### *5.1.2 3D plate with a stationary crack under tensile loading*

In this subsection, a 3D plate with a central crack is uniaxially loaded to test the ability of LPM to handle the discontinuous configurations. The size of the plate is 50mm*50mm*2mm with the through-thickness central crack of 10mm length. The loading direction is perpendicular to the crack face. The system undergoes a quasi-static loading of 120MPa.

To directly compare with the simulation results performed by FEM in [51], the simple cubic lattice structure is adopted to simulate the perfectly plastic behavior. The materials constants are listed in the following: the Young's modulus E is 200GPa, Poisson's ratio ν is 0.33, and yield stress $\sigma_y$ is 200MPa. The radius of the particle is 0.2mm, thus there are 5 layers of particles along the thickness direction and 126 layers of particles along loading direction. The total number of particles is 78750. And the loading area would therefore be 100.8mm$^2$.

For LPM simulations, the 120MPa load is subdivided into 30 steps with 4MPa for each step. The second layer of particles along the thickness direction is selected to output the field quantity. As shown in Figure 8, the yield particles (nodes), equivalent plastic strain, and the horizontal displacement component parallel to crack face are compared between LPM and FEM simulations. The LPM simulation results match well with the FEM counterparts.

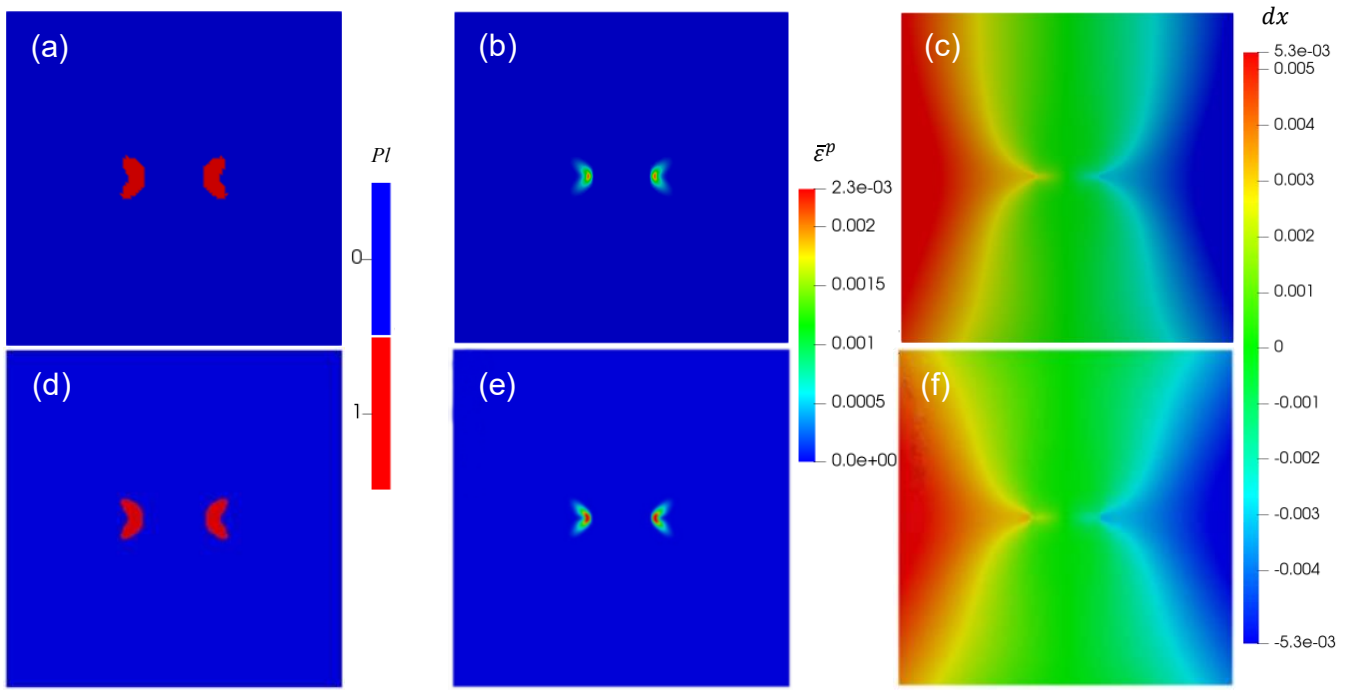

**Figure 8**. Distribution of yield particles, equivalent plastic strain, and horizontal displacement

field of 3D cracked plate for (a), (b), (c) for LPM and (d), (e), (f) for FEM (refer to [51])

## 5.2 Damage and fracture analysis using LPM

### 5.2.1 Brittle fracture analysis of 2D structures

The brittle fracture phenomena are studied in this section. Compared to the previous works of crack analysis using LPM [18], the iteration framework described in Section 4 is able to redistribute the displacement and force field during crack grow without having to adopt small loading steps. As shown in [50], the maximum number of broken bonds should be controlled as small as possible to get reasonable simulation results. For all the brittle fracture simulations, the maximum number of broken bonds per iteration is set as 1.

*Case I: Notched beam under three-point bending load*

In this example, the three-point bending test is simulated using 2D square lattice structure under plane stress assumption. The sample dimensions as well as the boundary conditions are sketched out in Figure 9 [52]. The material constants are such that: the Young's modulus E is 31.4GPa, Poisson's ratio $\nu$ is 0.4. The radius of the particle is 0.002m, thus there are 12460 particles in total for the system. The concentrated force $P$ is loaded by applying displacement boundary condition on the middle point of the beam's top boundary. The incremental displacement along x direction is $10^{-6}$ m.

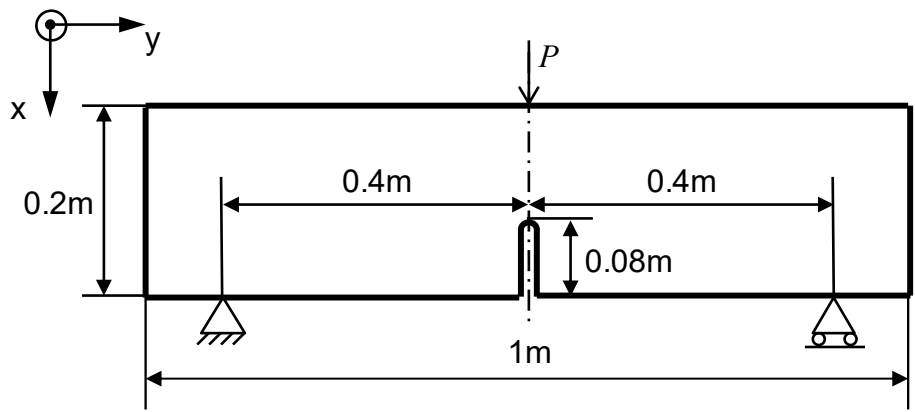

**Figure 9**. Dimensions and boundary conditions for the three-point bending test (refer to [52])

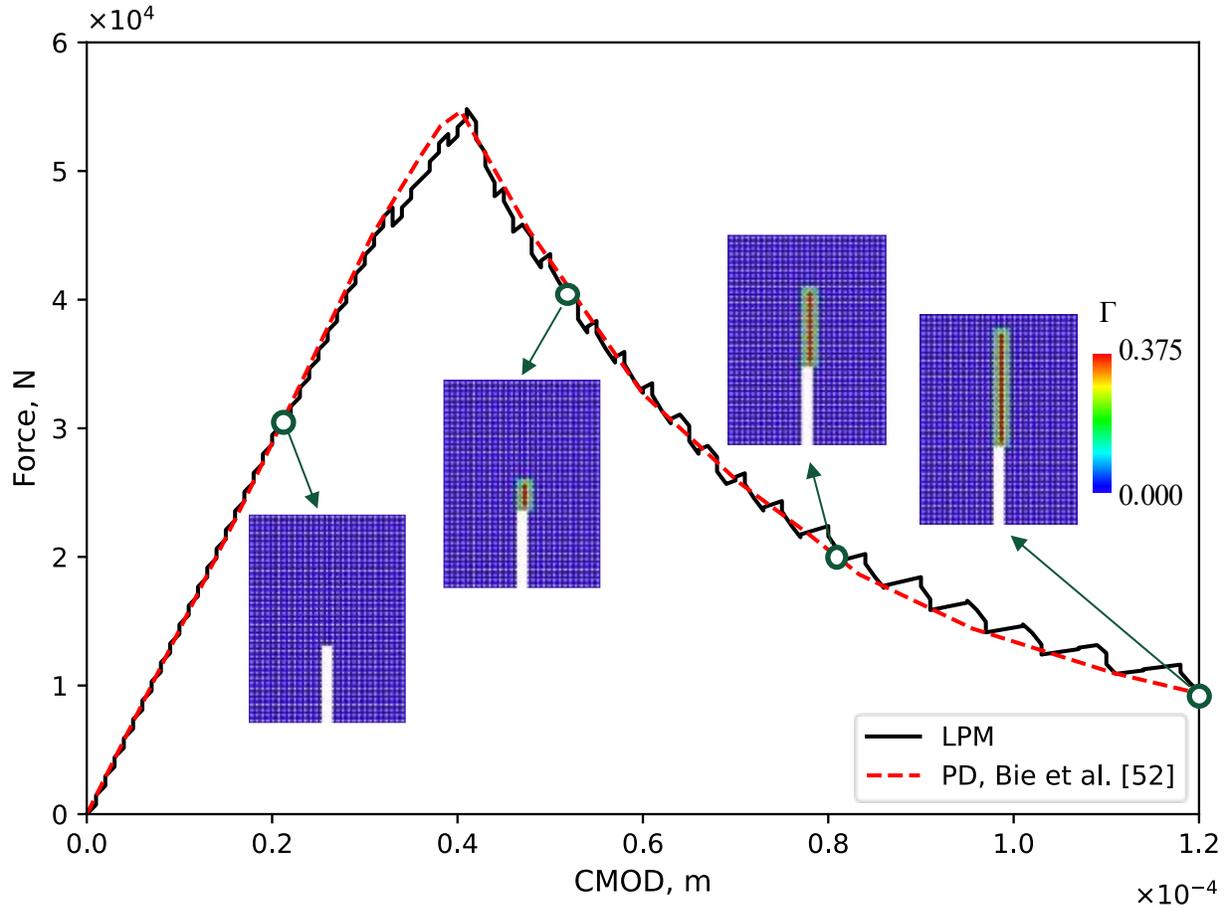

**Figure 10**. Load-CMOD curves obtained using the damage-enhanced lattice particle simulation as well as the deterioration ratio Γ distribution

After calibration, the critical bond strain value $\xi_{cr}$ is determined as $2.7 \times 10^{-4}$. The curves of the reaction force at loading point verses the crack mouth opening displacement (CMOD) are plotted in Figure 10. As shown in the figure, there is no deterioration occurs before the peak load. During the crack grow process after peak load, the load-carrying capability of the beam is weakened continuously. The simulated load-CMOD curve is in good agreement with that of Peridynamics (PD) by Bie et al. [52].

*Case II: Single edge notched plate under pure shear loading*

The example in this section is a shear test for a square specimen with notched edge under 2D plane strain assumption. Even though the problem setting is simple, it is a nontrivial simulation test that involves the asymmetric crack pattern [53]. As indicated in [18,38], the regular mesh geometry utilized by the LPM framework would result in a preferential direction during the crack propagation and may deviate from the real failure pattern. This lattice-dependence issue is demonstrated to be reduced if using more layers of neighbors [18]. However, there is no study that shows whether lattice shape has any effects on the crack growth direction.

The examples given below will utilize both the square and hexagon lattice structures with same particle radius (i.e., $3.2 \times 10^{-3}$mm) and number of neighbors (i.e., 2). The sample configuration and boundary condition are illustrated in Figure 11. The bottom layer of lattices is fixed in both x and y directions while the top layer is applied to displacement boundary condition. The incremental displacement for each loading step is $1.5 \times 10^{-4}$mm. After model calibrations, the critical bond strain value $\xi_{cr}$ is determined as $2.7 \times 10^{-2}$ for both lattice structures. The reaction force verses accumulated displacement curves are plotted in Figure 12.

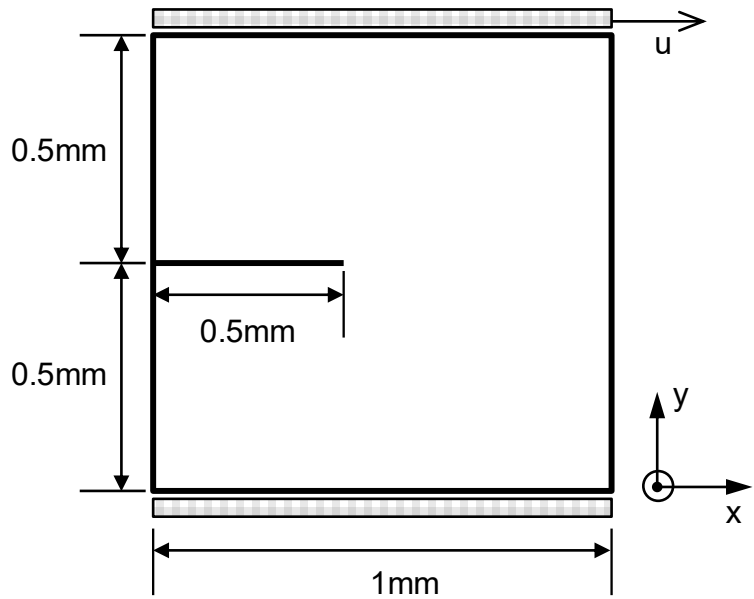

**Figure 11**. Geometry and loading conditions of single edge notched plate (refer to [50])

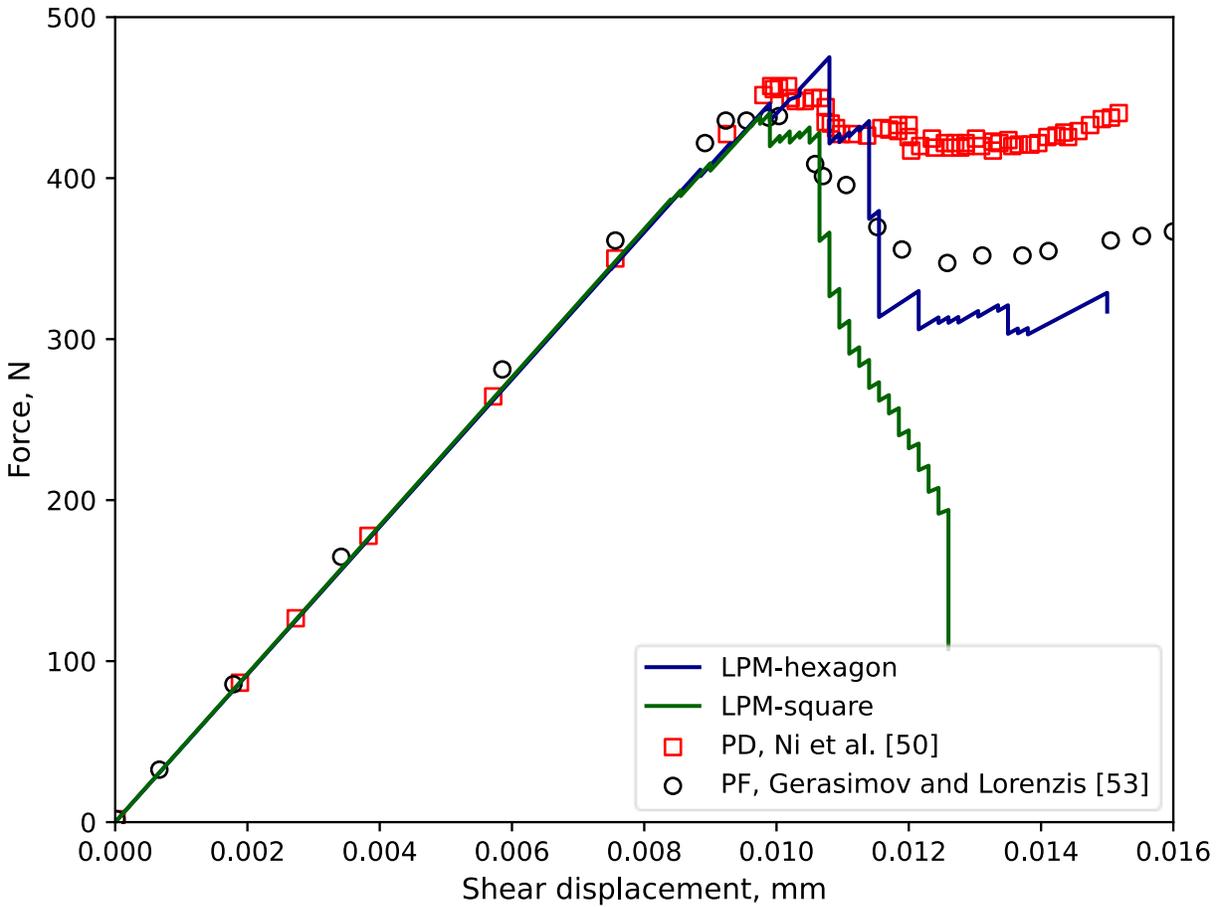

**Figure 12**. Load-displacement curves simulated using LPM for 2D hexagon and square lattice structures compared to results of PD [50] and PF [53]

In Figure 12, the simulation results of PD [50] and phase field (PF) models [53] are compared against to that of LPM. Both that using the 2D square and hexagon lattices can obtain reasonable peak force, which is around 440N. However, the force-displacement curve utilizing square lattice drops dramatically after reaching the peak loading. This indicates that the crack chooses to grow along a preferential direction in this regular packed lattice structure even 2 layers of neighbors are used. For the hexagon lattice, however, the curve stabilized and even increased a little bit after the initial drop after reaching the peak force. This is also in accordance to that of PD and PF simulation results.

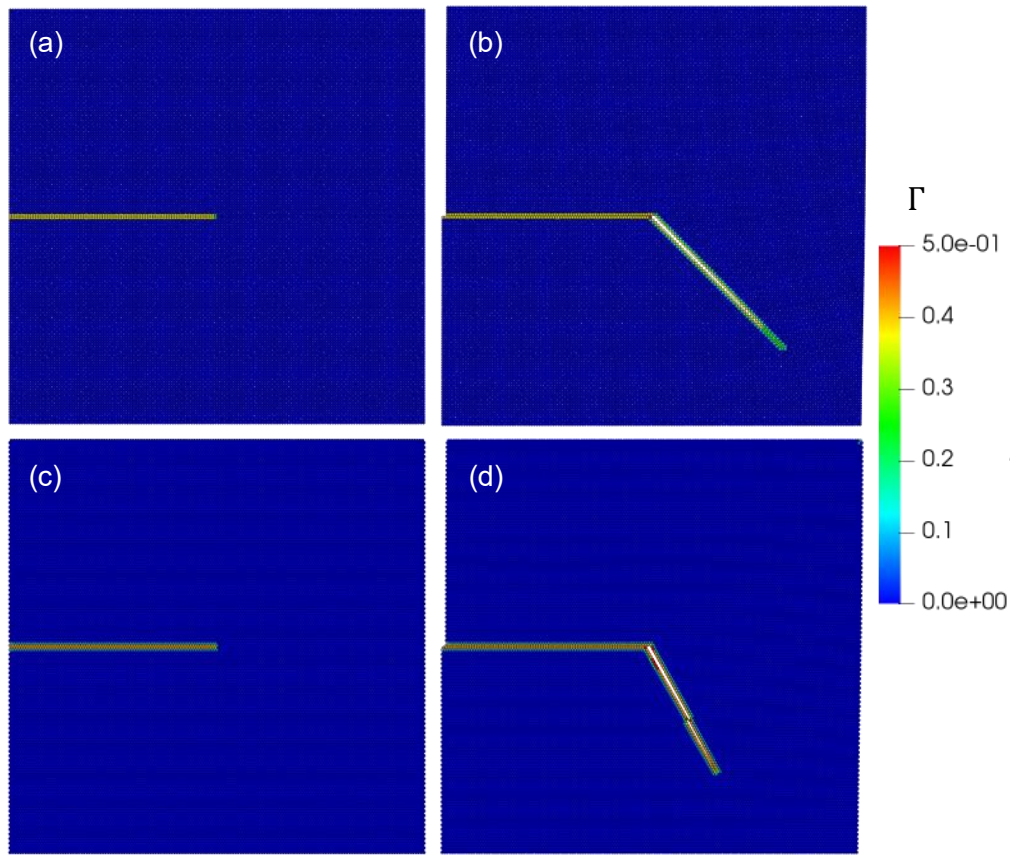

**Figure 13**. Deterioration ratio contours of (a) square lattice at initial time, (b) square lattice at shear displacement of about 0.012mm; (c) hexagon lattice at initial time, (d) hexagon lattice at shear displacement of about 0.012mm

The crack grow patterns for both lattice structures are plotted in Figure 13. It shows that the crack growth direction deviates each other for these two lattice structures. It is clear that the predicted pattern using hexagon lattice is closer to the results predicted by PF [53] or PD [50]. Looking back at the Table I, the number of neighbors of square lattice is 8 in total compared to 12 for hexagon lattices. This means the nonlocal effects are better handled by hexagon lattice structures by including more neighbors. However, the simulations using lattice structures with more neighbors usually costs much more than lattice structures with less neighbors. This is also a tradeoff in simulating various real problems.

## *5.2.2 Damage analysis of a 2D ductile plate*

In the current and the following sections, the particle-wise nonlocal damage accumulation rule described in Section 4.1 would be incorporated into the LPM framework. For the 2D ductile plate, the uniaxial tension loading is applied in a form of incremental displacements. As shown in Figure 14, the uniaxial displacement constraints are applied on the top and bottom layers of the system. The hexagon lattice structure with the assumption of plane strain is adopted in the following discussion. The number of broken bonds (particles) is not restricted for each iteration step.

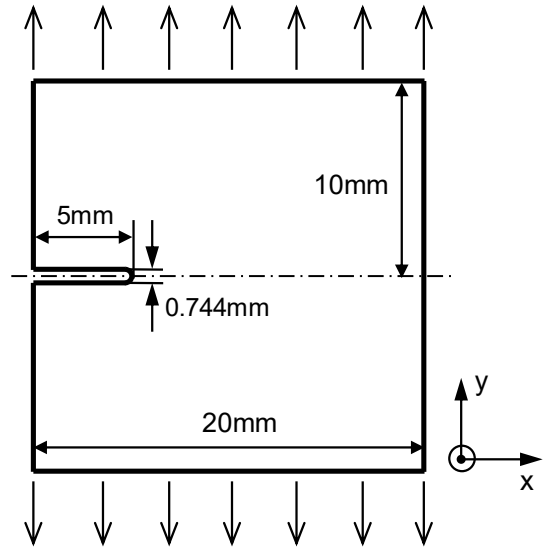

**Figure 14.** Geometry and boundary conditions for the square plate with notched edge

The materials are assumed as the same as the one in Section 5.1. The damage parameters are selected as $A=80$, $D_c=0.9$, and the nonlocal characteristic length $L=0.5$mm. The ductile fracture behavior might be directly simulated using elastoplastic constitutive relationships coupled with a global crack criterion, e.g., J-integral, like the work of Madenci and Oterkus [54]. However, this trivial treatment of ductile fracture simulation would result in the over-estimation of mechanical responses if no damage is applied (Figure 15(a)). However, the damage-enhanced LPM framework reproduced the nonlinear processes including elastic, plastic hardening, softening, and material deterioration occurred in real materials even though a linear hardening model is used.

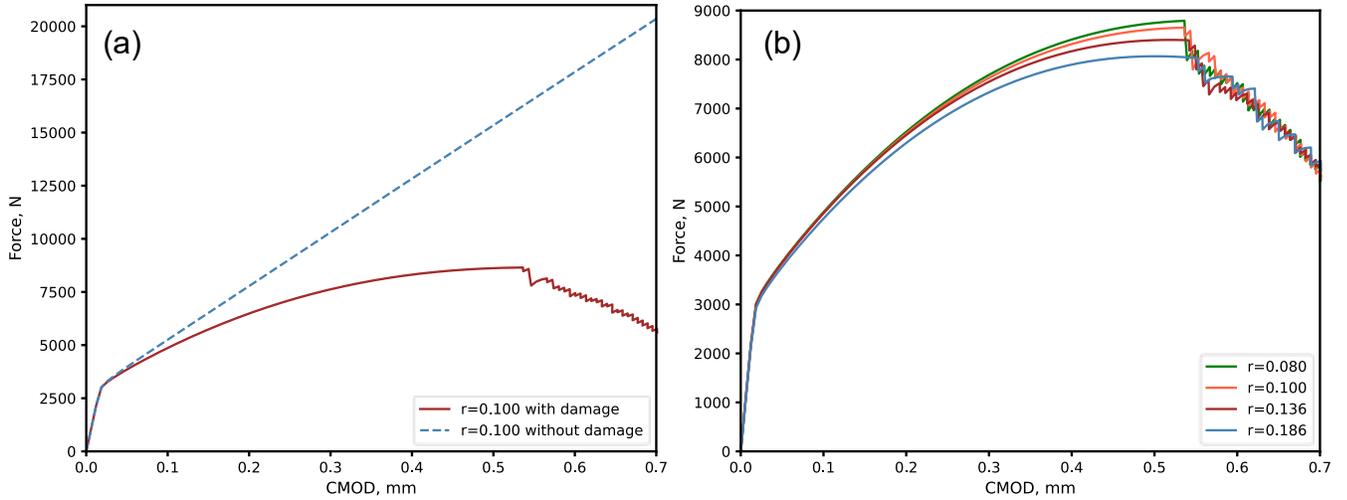

**Figure 15**. Comparisons of load-CMOD curves between (a) with/without damage when particle radius is 0.1mm; (b) lattice systems with different particle radius

To discuss the particle size-dependence issues, the load-CMOD curves for lattice systems with different particle radius are compared against each other, i.e., Figure 15(b). Even though it shows some deviations before the breakage occurs for the simulations with different particle radius, the load-CMOD curves in the post-peak stages are similar to each other. As shown in the figure, all of the bond breakages occur at the local maximum point of the load-CMOD curves. This indicates that the load-carrying ability of the plate cannot keep increasing because the damage variable is approaching 1. Therefore, the bond integrity parameter would approach zero and no longer have any load-carrying capabilities.

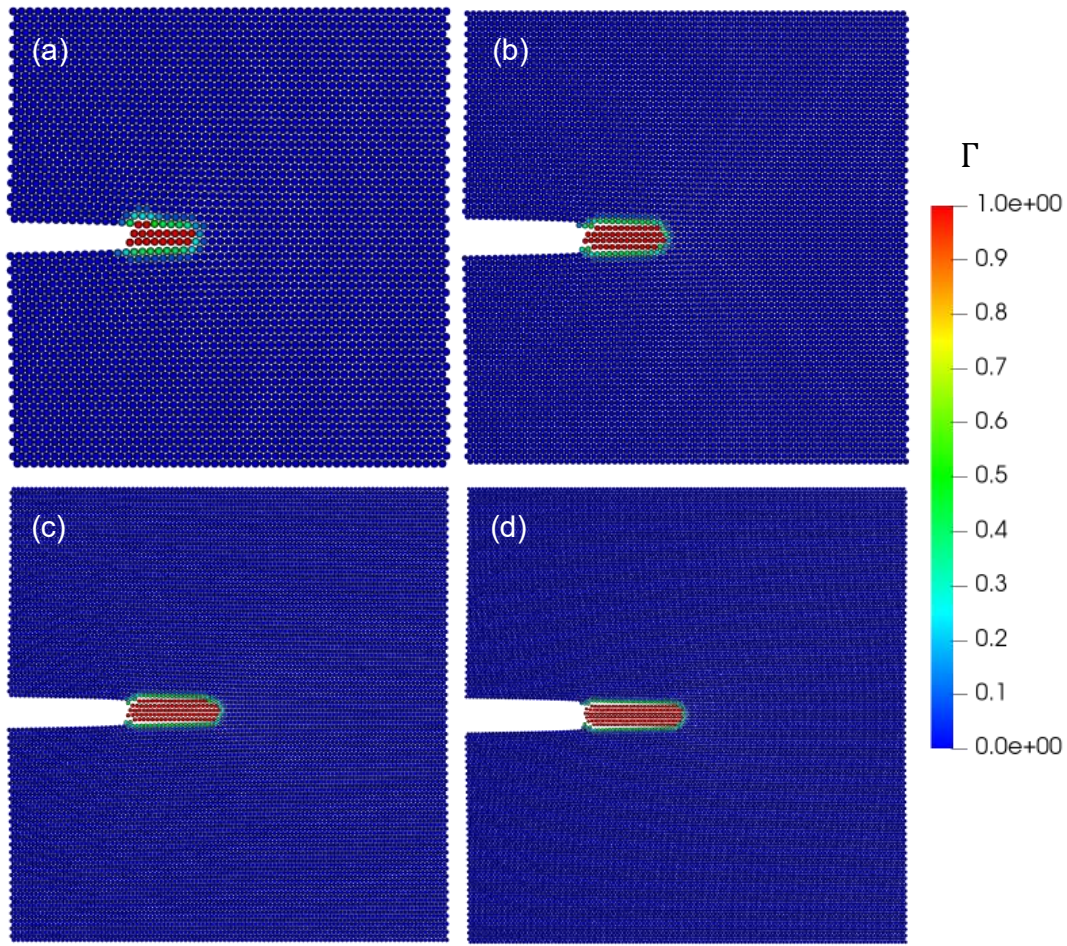

**Figure 16**. Distribution of deterioration ratio for ductile plate with particle radius of (a) 0.186mm, (b) 0.136mm, (c) 0.100mm, and (d) 0.080mm

The inaccuracy of LPM when using small particle densities may be the reason of getting different load-CMOD curves when varying particle densities [55], especially when there also involves a pre-exist crack. As is shown in Figure 15(b), all curves jump to the similar force level after bond breakage occurs. This phenomenon largely comes from the nonlocal damage formulation. Since the crack tip tends to have similar size and shape after bond breakage occurs. The distribution of the deterioration ratio $\Gamma$ for each particle is shown in Figure 16. It indicates that, no matter what size of the particle radius is, the deteriorated particles always form a stripe of similar width ahead of the pre-exist crack tip. The nonlocal damage contour also tends to be identical if reduce the particle radius, i.e., Figure 17. Above all, it can be concluded that the particle size-dependence can be reduced by including the nonlocal damage by keeping the characteristic length scale to be the same for each simulation.

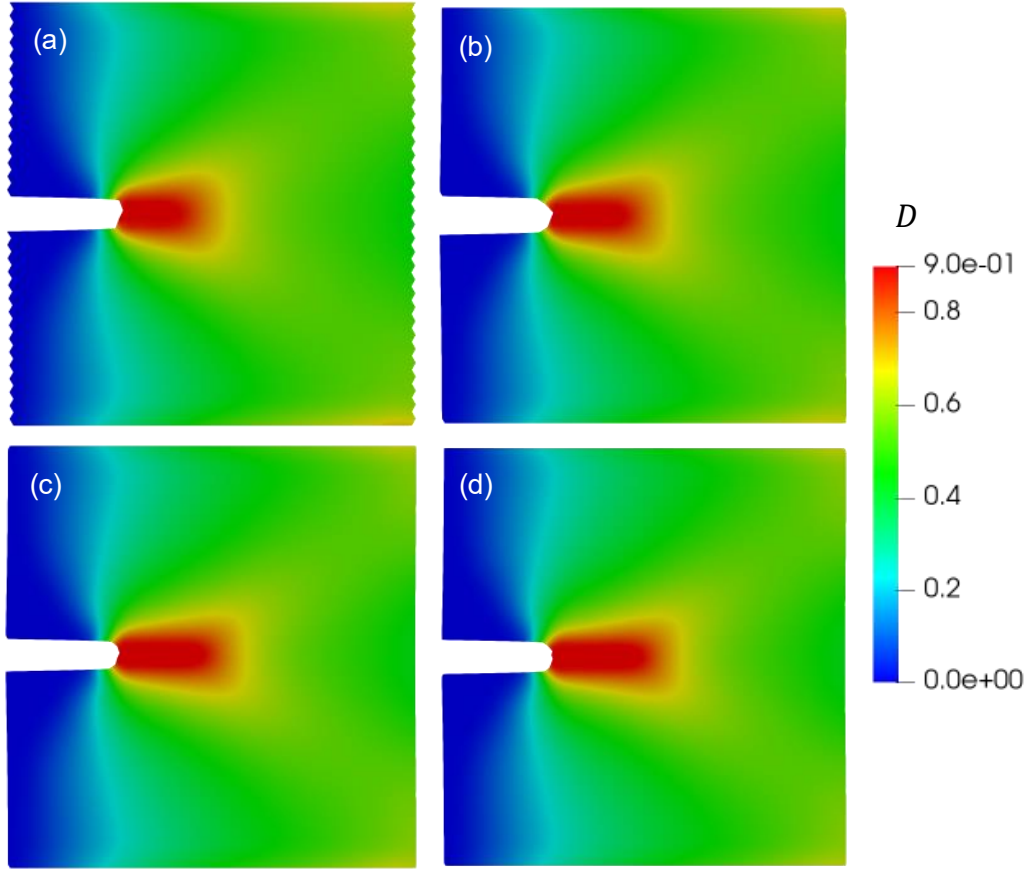

**Figure 17**. Nonlocal damage contour of the ductile plate with particle radius of (a) 0.186mm, (b) 0.136mm, (c) 0.100mm, and (d) 0.080mm

### *5.2.3 Fracture of a 3D compact tension plate specimen*

In this section, the damage-enhanced LPM framework is utilized to simulate the mechanical behavior of a compact tension (CT) test for titanium alloy. The original experiments are from the work of Ding et al. [56].The numerical simulation has also been done using the DLSM method by Zhao et al. [26]. The full 3D simple cubic lattice is adopted in this example, with the particle radius of 0.33mm. The corresponding material constants are taken from [56], which are Young's modulus

E is 115GPa, Poisson's ratio ν is 0.28, the initial yield stress $\sigma_y$ is 955MPa and the hardening modulus *H* is 2.4018MPa. The lattice configuration and geometry are shown in Figure 18. The particles near the loading point are applied to the incremental displacement boundary condition, which is 6.4× $10^{-3}$mm/step.

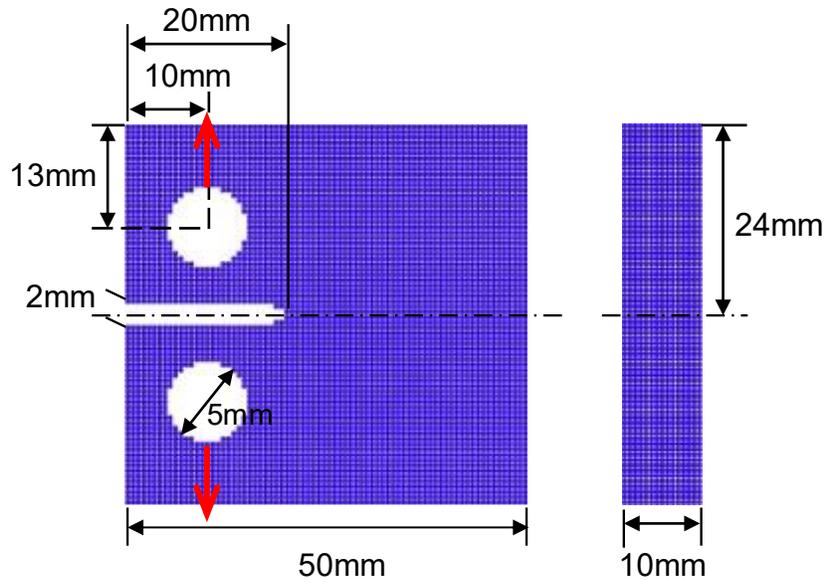

**Figure 18**. Geometry and boundary conditions of the 3D compact tension (CT) specimen

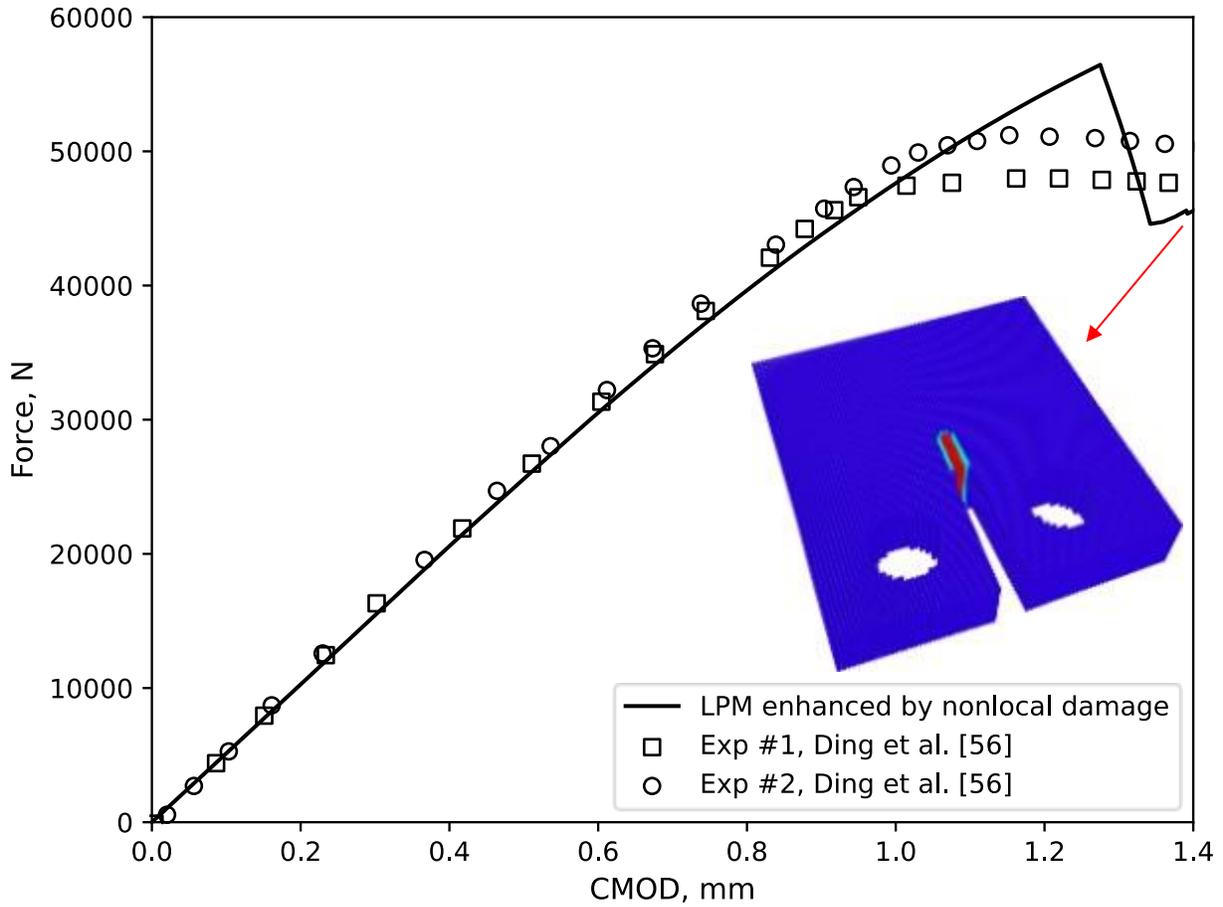

**Figure 19**. Load-CMOD curve of the CT test using damage-enhanced LPM framework

There are some particular cares that need to be taken to properly simulate the CT test. One of the most crucial assumptions is that the particles that higher than the top circle and particles that lower than the bottom circle need to have a high number of yield stress, i.e., $10^6$MPa. The reason of doing so is due to the fact of stress concentration, the lattice would distort much from its initial configuration and introduce undesired damage near the loading point after yielding occurs. After calibration, the damage parameters are determined as $A$=400, $D_c$ =0.85, and the nonlocal characteristic length $L$=0.6mm. As shown in Figure 19, the fracture pattern as well as the load-CMOD curve are consistent with the experimental observation.

## 6. Conclusions

LPM is a powerful method to simulate the solid mechanics problems involving discontinuous stationary and growing cracks. The return-mapping framework based on the iterative solution procedure was proposed in this paper as an improvement over the previous incremental updating J2 plasticity model. In addition, the bond-wise and particle-wise nonlocal damage models were incorporated into the nonlocal LPM framework. Several benchmark examples were tested using the damage-enhanced elastoplastic simulation framework. The following conclusions can be made:

1) By exploring the interchangeability between particle-wise and bond-wise properties, the continuum elastoplastic constitutive relationship that based on tensor quantities can be implemented into LPM using the return mapping algorithm.

2) The distortional energy-based incremental updating algorithm is found to always underestimate the yielding point. However, the LPM framework using tensor-based return-mapping algorithm with the iterative solution procedure is more accurate, stable and general compared to the energy-based method.

3) The crack propagation algorithm of proposed LPM framework can force the system to be in equilibrium and would get more reasonable crack growth and macroscopic mechanical behavior.

4) With the same number of neighbor layers, the lattices with denser packing are more likely to be able to get reasonable crack grow pattern.

5) By introducing the nonlocal ductile damage accumulation rule, the particle size-dependency issue of damage modelling can be reduced. The deterioration contours as well as the macroscopic mechanical responses are expected to be converged for thinner particle size.

The proposed method is consistent with the classic J2 plasticity theory and thus paved a way to consider general material non-linearity and damage process in the LPM simulation. In the future, the randomness would be considered to formulate the nonlocal damage model to reproduce the heterogeneous nature of real structures. In this case, more complicated mechanical behaviors may be simulated, such as fatigue behaviors. The nonlocal damage-enhanced LPM framework could also be extended into the large deformation regime to more accurately capture the softening behavior near damaged areas.

## Acknowledgements

The research is partially supported by fund from NAVAIR through subcontract from Technical Data Analysis, Inc (TDA) (contract No. N68-335-18-C-0748, program manager: Krishan Goel). The support is greatly appreciated. The inspiring discussions held with Dr. Haoyang Wei and Professor Hailong Chen are gratefully acknowledged.

# References

[1] A. Pineau, A.A. Benzerga, T. Pardoen, Failure of metals I: Brittle and ductile fracture, Acta Mater. 107 (2016) 424–483. https://doi.org/10.1016/j.actamat.2015.12.034.

[2] U. Yolum, A. Taştan, M.A. Güler, A Peridynamic Model for Ductile Fracture of Moderately Thick Plates, Procedia Struct. Integr. 2 (2016) 3713–3720. https://doi.org/10.1016/j.prostr.2016.06.461.

[3] M. Ambati, T. Gerasimov, L. De Lorenzis, Phase-field modeling of ductile fracture, Comput. Mech. 55 (2015) 1017–1040. https://doi.org/10.1007/s00466-015-1151-4.

[4] A.L. Gurson, Continuum Theory of Ductile Rupture by Void Nucleation and Growth : Part 1 — Yield Criteria and Flow Rules for Porous Ductile Media, J. Eng. Mater. Technol. 99 (1977) 2–15.

[5] V. Tvergaard, A. Needleman, Analysis of the cup-cone fracture in a round tensile bar, Acta Metall. 32 (1984) 157–169.

[6] L.M. Kachanov, Introduction to continuum damage mechanics, Kluwer Academic Publishers, Dordrecht, 1986.

[7] J. Lemaitre, A course on damage mechanics, Springer-Verlag Berlin Heidelberg, 1996.

[8] J.L. Chaboche, Continuous damage mechanics - A tool to describe phenomena before crack initiation, Nucl. Eng. Des. 64 (1981) 233–247. https://doi.org/10.1016/0029-5493(81)90007-8.

[9] L. De Lorenzis, D. Fernando, J.G. Teng, Coupled mixed-mode cohesive zone modeling of interfacial debonding in simply supported plated beams, Int. J. Solids Struct. 50 (2013) 2477–2494. https://doi.org/10.1016/j.ijsolstr.2013.03.035.

[10] T. Belytschko, W.K. Liu, B. Moran, K. Elkhodary, Nonlinear finite elements for continua and structures, 2014. https://www.wiley.com/en-us/Nonlinear+Finite+Elements+for+Continua+and+Structures%2C+2nd+Edition-p-9781118632703 (accessed September 19, 2020).

[11] N. Moës, J. Dolbow, T. Belytschko, A finite element method for crack growth without remeshing, Int. J. Numer. Methods Eng. 46 (1999) 131–150. https://doi.org/10.1002/(SICI)1097-0207(19990910)46:1<131::AID-NME726>3.0.CO;2-J.

[12] G.D. Nguyen, A.M. Korsunsky, J.P.H. Belnoue, A nonlocal coupled damage-plasticity model for the analysis of ductile failure, Int. J. Plast. 64 (2015) 56–75.

https://doi.org/10.1016/j.ijplas.2014.08.001.

[13] J. Besson, Continuum models of ductile fracture: A review, Int. J. Damage Mech. 19 (2010) 3–52. https://doi.org/10.1177/1056789509103482.

[14] C. Kuhn, T. Noll, R. Müller, On phase field modeling of ductile fracture, GAMM Mitteilungen. 39 (2016) 35–54. https://doi.org/10.1002/gamm.201610003.

[15] C. Miehe, F. Aldakheel, A. Raina, Phase field modeling of ductile fracture at finite strains: A variational gradient-extended plasticity-damage theory, Int. J. Plast. 84 (2016) 1–32. https://doi.org/10.1016/j.ijplas.2016.04.011.

[16] S.A. Silling, R.B. Lehoucq, Peridynamic Theory of Solid Mechanics, Adv. Appl. Mech. 44 (2010) 73–168. https://doi.org/10.1016/S0065-2156(10)44002-8.

[17] S.A. Silling, M. Epton, O. Weckner, J. Xu, E. Askari, Peridynamic states and constitutive modeling, J. Elast. 88 (2007) 151–184. https://doi.org/10.1007/s10659-007-9125-1.

[18] H. Chen, E. Lin, Y. Jiao, Y. Liu, A generalized 2D non-local lattice spring model for fracture simulation, Comput. Mech. 54 (2014) 1541–1558.

[19] A. Hrennikoff, Solution of Problems of Elasticity by the Framework Method, J. Appl. Mech. 8 (1941) A169–A175. https://doi.org/10.1115/1.4009129.

[20] H. Gao, P. Klein, Numerical simulation of crack growth in an isotropic solid with randomized internal cohesive bonds, J. Mech. Phys. Solids. 46 (1998) 187–218. https://doi.org/10.1016/S0022-5096(97)00047-1.

[21] M. Ostoja-Starzewski, Lattice models in micromechanics, Appl. Mech. Rev. 55 (2002) 35–59. https://doi.org/10.1115/1.1432990.

[22] Z. Zhang, Y. Chen, Modeling nonlinear elastic solid with correlated lattice bond cell for dynamic fracture simulation, Comput. Methods Appl. Mech. Eng. 279 (2014) 325–347. https://doi.org/10.1016/j.cma.2014.06.036.

[23] G.F. Zhao, Developing a four-dimensional lattice spring model for mechanical responses of solids, Comput. Methods Appl. Mech. Eng. 315 (2017) 881–895. https://doi.org/10.1016/j.cma.2016.11.034.

[24] G. Zhao, J. Fang, J. Zhao, A 3D distinct lattice spring model for elasticity and dynamic failure, Int. J. Numer. Anal. Methods Geomech. 35 (2011) 859–885. https://doi.org/10.1002/nag.

[25] Z. Pan, R. Ma, D. Wang, A. Chen, A review of lattice type model in fracture mechanics: theory, applications, and perspectives, Eng. Fract. Mech. 190 (2018) 382–409. https://doi.org/10.1016/j.engfracmech.2017.12.037.

A Nonlocal Damage-enhanced Lattice Particle Model for Ductile Fracture Analysis[26] G.F. Zhao, J. Lian, A. Russell, N. Khalili, Implementation of a modified Drucker–Prager model in the lattice spring model for plasticity and fracture, Comput. Geotech. 107 (2019) 97–109. https://doi.org/10.1016/j.compgeo.2018.11.021.

[27] H. Chen, E. Lin, Y. Liu, A novel Volume-Compensated Particle method for 2D elasticity and plasticity analysis, Int. J. Solids Struct. 51 (2014) 1819–1833. http://dx.doi.org/10.1016/j.ijsolstr.2014.01.025.

[28] H. Wei, H. Chen, Y. Liu, A nonlocal lattice particle model for J2 plasticity, Int. J. Numer. Methods Eng. 121 (2020) 5469–5489. https://doi.org/10.1002/nme.6446.

[29] M.R. Tupek, J.J. Rimoli, R. Radovitzky, An approach for incorporating classical continuum damage models in state-based peridynamics, Comput. Methods Appl. Mech. Eng. 263 (2013) 20–26. https://doi.org/10.1016/j.cma.2013.04.012.

[30] R. Bargellini, J. Besson, E. Lorentz, S. Michel-Ponnelle, A non-local finite element based on volumetric strain gradient: Application to ductile fracture, Comput. Mater. Sci. 45 (2009) 762–767. https://doi.org/10.1016/j.commatsci.2008.09.020.

[31] V. Tvergaard, A. Needleman, Effects of nonlocal damage in porous plastic solids, Int. J. Solids Struct. 32 (1995) 1063–1077. https://doi.org/10.1016/0020-7683(94)00185-Y.

[32] H. Chen, Y. Xu, Y. Jiao, Y. Liu, A novel discrete computational tool for microstructure-sensitive mechanical analysis of composite materials, Mater. Sci. Eng. A. 659 (2016) 234–241. https://doi.org/10.1016/j.msea.2016.02.063.

[33] H. Chen, Y. Liu, A non-local 3D lattice particle framework for elastic solids, Int. J. Solids Struct. 81 (2016) 411–420. http://dx.doi.org/10.1016/j.ijsolstr.2015.12.026.

[34] H. Chen, A Novel Nonlocal Lattice Particle Framework for Modeling of Solids, Arizona State University, 2015.

[35] H. Chen, Y. Liu, Y. Jiao, C. Meng, Modeling elasticity of cubic polycrystals using a nonlocal lattice particle method. Manuscript in preparation, (2020).

[36] B. Kilic, E. Madenci, An adaptive dynamic relaxation method for quasi-static simulations using the peridynamic theory, Theor. Appl. Fract. Mech. 53 (2010) 194–204. https://doi.org/10.1016/j.tafmec.2010.08.001.

[37] B. Liu, Y. Huang, H. Jiang, S. Qu, K.C. Hwang, The atomic-scale finite element method, Comput. Methods Appl. Mech. Eng. 193 (2004) 1849–1864.

[38] E. Lin, H. Chen, Y. Liu, Finite element implementation of a non-local particle method for elasticity and fracture analysis, Finite Elem. Anal. Des. 93 (2015) 1–11. http://dx.doi.org/10.1016/j.finel.2014.08.008.
46

[39] M.D. Brothers, J.T. Foster, H.R. Millwater, A comparison of different methods for calculating tangent-stiffness matrices in a massively parallel computational peridynamics code, Comput. Methods Appl. Mech. Eng. 279 (2014) 247–267. https://doi.org/10.1016/j.cma.2014.06.034.

[40] H. Chen, Constructing continuum-like measures based on a nonlocal lattice particle model: Deformation gradient, strain and stress tensors, Int. J. Solids Struct. 169 (2019) 177–186. https://doi.org/10.1016/j.ijsolstr.2019.04.014.

[41] E. de S. Neto, D. Peric, D. Owen, Computational methods for plasticity: theory and applications, 2011.

[42] J.C. Simo, T.J. Hughes, Computational inelasticity. Vol. 7., Springer Science & Business Media, 2006.

[43] C. Meng, H. Wei, H. Chen, Y. Liu, Modeling plasticity of cubic crystals using a nonlocal lattice particle method, Comput. Methods Appl. Mech. Eng. 385 (2021) 114069. https://doi.org/10.1016/j.cma.2021.114069.

[44] H. Chen, Y. Jiao, Y. Liu, Investigating the microstructural effect on elastic and fracture behavior of polycrystals using a nonlocal lattice particle model, Mater. Sci. Eng. A. 631 (2015) 173–180. https://doi.org/10.1016/j.msea.2015.02.046.

[45] J.T. Foster, S.A. Silling, W. Chen, An energy based failure criterion for use with peridynamic states, Int. J. Multiscale Comput. Eng. 9 (2011) 675–687. https://doi.org/10.1615/IntJMultCompEng.2011002407.

[46] M. Behzadinasab, J.T. Foster, A semi-Lagrangian constitutive correspondence framework for peridynamics, J. Mech. Phys. Solids. 137 (2020) 103862. https://doi.org/10.1016/j.jmps.2019.103862.

[47] H. Li, M.W. Fu, J. Lu, H. Yang, Ductile fracture : Experiments and computations, Int. J. Plast. 27 (2011) 147–180. https://doi.org/10.1016/j.ijplas.2010.04.001.

[48] K. Enakoutsa, J.B. Leblond, G. Perrin, Numerical implementation and assessment of a phenomenological nonlocal model of ductile rupture, Comput. Methods Appl. Mech. Eng. 196 (2007) 1946–1957. https://doi.org/10.1016/j.cma.2006.10.003.

[49] M. Oyane, Criteria of ductile fracture strain, Bull. JSME. 15 (1972) 1507–1513.

[50] T. Ni, M. Zaccariotto, Q.Z. Zhu, U. Galvanetto, Static solution of crack propagation problems in Peridynamics, Comput. Methods Appl. Mech. Eng. 346 (2019) 126–151. https://doi.org/10.1016/j.cma.2018.11.028.

[51] M. Asgari, M.A. Kouchakzadeh, An equivalent von Mises stress and corresponding

equivalent plastic strain for elastic–plastic ordinary peridynamics, Meccanica. 54 (2019) 1001–1014. https://doi.org/10.1007/s11012-019-00975-8.

[52] Y. Bie, S. Li, X. Hu, X. Cui, An implicit dual-based approach to couple peridynamics with classical continuum mechanics, Int. J. Numer. Methods Eng. 120 (2019) 1349–1379. https://doi.org/10.1002/nme.6182.

[53] T. Gerasimov, L. De Lorenzis, A line search assisted monolithic approach for phase-field computing of brittle fracture, Comput. Methods Appl. Mech. Eng. 312 (2016) 276–303. https://doi.org/10.1016/j.cma.2015.12.017.

[54] E. Madenci, S. Oterkus, Ordinary state-based peridynamics for plastic deformation according to von Mises yield criteria with isotropic hardening, J. Mech. Phys. Solids. 86 (2016) 192–219. https://doi.org/10.1016/j.jmps.2015.09.016.

[55] H. Chen, Y. Jiao, Y. Liu, A nonlocal lattice particle model for fracture simulation of anisotropic materials, Compos. Part B Eng. 90 (2016) 141–151. https://doi.org/10.1016/j.compositesb.2015.12.028.

[56] J. Ding, Z. Zhang, F. Yang, Y. Zhao, X. Ge, Plastic fracture simulation by using discretized virtual internal bond, Eng. Fract. Mech. 178 (2017) 169–183. https://doi.org/10.1016/j.engfracmech.2017.04.032.